\documentclass[12pt]{article}
\usepackage{fullpage,amsmath,amsfonts,graphicx,amsthm,url}
\usepackage{epstopdf}
\usepackage{natbib}

\def\H{{\bf{H}}}

\def\y{{\bf{y}}}
\def\z{{\bf{z}}}

\def\X{{\bf{X}}}
\def\Z{{\bf{Z}}}
\def\P{{\bf{P}}}

\def\S{{\bf{S}}}
\def\C{{\bf{C}}}

\def\A{{\bf{A}}}

\def\Gam{{\bf{\Gamma}}}

\def\L{{\bm{L}}}

\def\xobs{{\bf{\y_{\rm{o}}}}}
\def\zobs{{\bf{\y_{\rm{o}}}}}

\def\Id{{\bf{I}}}
\def\tobs{t_{\rm{obs}}}
\def\robs{{\bf{r}}_{\rm{o}}}
\def\Robs{{\bf{R}}_{\rm{o}}}
\def\RobsI{{\bf{R}}^{-1}_{\rm{o}}}

\def\PfI{{\bf{P}}^{-1}_{\rm{f}}}

\def\KR{{\bf{K}}_{\rm{o}}}

\def\PfI{{\bf{P}}_f^{-1}}

\def\d{\mathrm d}

\def\O{\mathcal{O}}
\def\L{\mathcal{L}}
\def\eps{\varepsilon}

\newcommand{\myabstract}{
A deterministic multiscale toy model is studied in which a chaotic fast subsystem triggers rare transitions between slow regimes, akin to weather or climate regimes. Using homogenization techniques, a reduced stochastic parametrization model is derived for the slow dynamics. The reliability of this reduced climate model in reproducing the statistics of the slow dynamics of the full deterministic model for finite values of the time scale separation is numerically established. The statistics however is sensitive to uncertainties in the parameters of the stochastic model.\\ It is investigated whether the stochastic climate model can be beneficial as a forecast model in an ensemble data assimilation setting, in particular in the realistic setting when observations are only available for the slow variables. The main result is that reduced stochastic models can indeed improve the analysis skill, when used as forecast models instead of the perfect full deterministic model. The stochastic climate model is far superior at detecting transitions between regimes. The observation intervals for which skill improvement can be obtained are related to the characteristic time scales involved. The reason why stochastic climate models are capable of producing superior skill in an ensemble setting is due to the finite ensemble size; ensembles obtained from the perfect deterministic forecast model lacks sufficient spread even for moderate ensemble sizes. Stochastic climate models provide a natural way to provide sufficient ensemble spread to detect transitions between regimes. This is corroborated with numerical simulations. The conclusion is that stochastic parametrizations are attractive for data assimilation despite their sensitivity to uncertainties in the parameters.}

\begin{document}

\title{Data assimilation in slow-fast systems using homogenized climate models}

\author{\textsc{Lewis Mitchell}
				\thanks{E-mail: lewism@maths.usyd.edu.au}
				\quad\textsc{and Georg A. Gottwald}
				\thanks{E-mail: georg.gottwald@sydney.edu.au}
				\\
\textit{\footnotesize{School of Mathematics and Statistics, University of Sydney, NSW 2006, Australia.}}
}
%

\maketitle

{
\begin{abstract}
\myabstract
\end{abstract}
\newpage
}


\section{Introduction}

An area of broad research in the atmospheric sciences in recent times is how to most effectively parametrize subgrid-scale processes. Since the pioneering papers by \cite{Leith75} and \cite{Hasselmann76}, the stochastic parametrization of processes which cannot be spatially and/or temporally resolved have recently gained popularity across disciplines in the atmospheric, oceanographic and climate sciences \citep{Palmer01}. Applications range from the study of atmospheric low-frequency variability \citep{Franzke05,FranzkeMajda06}, deep-convection and cloud modelling in GCM's \citep{LinNeelin00,LinNeelin02,PlantCraig08}, decadal climate changes such as El Ni\~no \citep{Kleeman08} to paleoclimatic modelling \citep{Ditlevsen99, KwasniokLohmann09}. We refer the reader to the books edited by \cite{Imkeller} and by \cite{PalmerWilliams} which contain excellent overviews of current trends in stochastic climate modelling.\\

\noindent
There exists a plethora of different methods to construct stochastic subgrid-scale parametrizations, including phenomenological approaches such as randomization of existing deterministic parametrization schemes (e.g. \cite{Buizza99}), energetic backscattering (e.g. \cite{Frederiksen97,Shutts05}), data driven techniques such as Markov chains (e.g. \cite{CrommelinVandenEijnden08}), and systematic approaches using stochastic homogenization (e.g. \cite{MTV99,Majda03}). The specific functional forms and parameter values of the respective parametrization schemes can be heuristically postulated on grounds of the particular physics involved (e.g. \cite{LinNeelin00}), or estimated using time series analysis (e.g. \cite{Wilks05}). In the case of multiscale dynamics, however, the parameters can be systematically derived using averaging and homogenization techniques (e.g. \cite{MTV99,Majdaetal01,Majdaetal08}). Our work will be concerned with the latter approach. The general theory of stochastic averaging and homogenization for multiscale dynamical systems goes back to the seminal papers of \cite{Khasminsky66}, \cite{Kurtz73} and \cite{Papanicolaou76}. Starting with \cite{MTV99} these mathematically rigorous  ideas have recently been applied in the atmospheric context by \cite{Franzke05} and \cite{FranzkeMajda06}.\\ 

\noindent
We will introduce a simple deterministic multiscale toy model in which a slow degree of freedom with multiple stable states, resembling slow weather or climate regimes (eg. \cite{LegrasGhil85,Crommelin03,Crommelin04,BranstatorBerner05} and \cite{Ditlevsen99, KwasniokLohmann09}), is driven by a fast chaotic subsystem which also involves metastable regimes. Although simple, this model involves core phenomena found in realistic applications such as slow and fast metastable states as caricatures of climate and weather regimes respectively, and rare transitions between them. The model is amenable to the theory of homogenization (for an excellent review see \cite{Givonetal04} and \cite{PavliotisStuart}), and we will derive a reduced stochastic climate model describing the slow dynamics of the full higher-dimensional model. Whereas for non-systematic stochastic parametrizations the results are sensitive to details of the assumed processes, we will verify here that the homogenized reduced model faithfully represents the slow dynamics of the full system even for moderate time scale separation. We have chosen a system amenable to homogenization precisely for the reason that the issue of the appropriate choice of the stochastic parametrization does not arise and the validity of the parametrization is guaranteed by rigorous theorems (at least in the limit of large time scale separation).\\

\noindent
Our main focus here is how well homogenized stochastic climate models perform when used as forecast models in the context of ensemble data assimilation. In data assimilation one attempts to find an optimal estimate of the state of a system combining information of noisy observations and a numerical forecast model exhibiting model error \citep{Kalnay}. This procedure is complicated by the nonlinear chaotic dynamics of the system as well as the impossibility of observing the entire system at any one time. On larger time scales, it is often only possible to adequately observe the slow, large scale degrees of freedom of the system, while the fast, small scale degrees of freedom in general remain unobservable. We consider one of the state-of-the art data assimilation methods called the ensemble Kalman filter (EnKF) \citep{Evensen94,Evensen}. In such a filter one evolves an ensemble of state estimates forward in time using a full nonlinear forecast model, and then estimates the forecast (or background) mean and its associated error variance from the ensemble. Together with observations of the system, this covariance is used in a least squares minimization problem to find an optimal state estimate called the analysis, along with an estimate of the associated analysis error covariance. This filter is optimal for linear systems with Gaussian errors, assumptions which are generically not consistent with real world systems. Besides their ease of implementation, the attractive feature of ensemble filters is that the forecast error covariance is estimated using the full nonlinear forecast model.\\

\noindent
Ensemble based Kalman filters, however, suffer from the problem of sampling errors due to the insufficient ensemble size. These errors usually underestimate the error covariances which may ultimately lead to filter divergence, when the filter trusts the forecast and ignores the information given by the observations. 
To avoid filter divergence the concept of covariance inflation was introduced whereby the prior forecast error covariance is increased by an inflation factor \citep{AndersonAnderson99}. This is usually done globally and involves careful and computationally expensive tuning of the inflation factor (for recent methods on adaptive estimation of the inflation factor from the innovation statistics see \cite{Anderson07,Anderson09,Li09}).\\

\noindent
We will address here the following questions: Under what circumstances can reduced stochastic climate models improve the skill of an ensemble based data assimilation scheme if used as forecast models? Furthermore, if they do, why so?\\ 

\noindent
\cite{HarlimMajda08, HarlimMajda10} studied ensemble filtering on the Lorenz-96 model \citep{Lorenz96} in the fully turbulent regime, where they found skill improvement for a stochastic climate forecast model which was constructed by radically replacing all nonlinear terms by linear stochastic processes whose statistics are estimated by fitting the model to the climatological variance, the decorrelation time and the observation time correlation. Here we study the performance of systematically derived climate models, however our results will also shed light on skill improvements seen in other ensemble based stochastic filtering methods.\\

\noindent
Stochastic climate models have often been found to not reproduce the autocorrelation function of the full deterministic system well, in particular for small time scale separation (see for example \cite{Franzke05,FranzkeMajda06}). Here we will see that this inaccuracy in faithfully reproducing the statistics of the slow dynamics does not preclude stochastic reduced models from being beneficial in data assimilation. We will identify a range of observation intervals in which the reduced stochastic climate model actually outperforms the full deterministic model. The range of observation intervals for which skill improvement is observed will be related to the characteristic time scales of the system. In particular we find that the observation interval has to be larger than the typical time taken for the slow state to switch regimes, and smaller than the decay rate of the autocorrelation function of the slow variable. Hence, stochastic climate models will be beneficial when studying weather or climate systems on time scales which resolve regime switches.\\

\noindent
We find that although climate models with appropriately determined drift and diffusion terms faithfully reproduce the slow dynamics of a multiscale deterministic model, the statistics of the climate model is rather sensitive to uncertainties in the drift and diffusion terms. A central result will be that their superior performance in data assimilation is not linked to their accurate approximation of the statistical slow behaviour but rather to a controlled increase of the ensemble spread associated with their inherent dynamic stochasticity, with an effect similar to inflating covariances or increasing ensemble size. Hence optimal performance of stochastic reduced climate models will be achieved when the diffusion coefficient is larger than the value which produces the closest match to the full deterministic dynamics.\\ 

\noindent
The remainder of the paper is organised as follows: In Section \ref{sec:model} we introduce the model and derive the reduced stochastic climate model for the slow dynamics using homogenization techniques. Estimation of the drift and diffusion terms in the climate model is performed in Section \ref{sec:paraest}. In Section \ref{sec:timescales} we examine some of the relevant characteristic time scales of the dynamics. In Section \ref{sec:sensitivity} we then study the sensitivity of these time scales to uncertainties in the time scale separation parameter as well as in the diffusion and the drift term of the climate model. After a brief overview of ensemble Kalman filtering in Section \ref{sec:EnKF} we present numerical results showing the skill improvement in using the reduced climate model over the full deterministic model in an ensemble Kalman filter in Section \ref{sec:results}. We conclude with a discussion in Section \ref{sec:discussion}.


\section{The model}
\label{sec:model}

We consider the following deterministic system with slow-fast time scale separation proposed in \cite{Givonetal04}
\begin{align}
	\frac{dx}{dt} &= x-x^3 +\frac{4}{90\varepsilon}y_2 \label{eqn:x}\\
	\frac{dy_1}{dt}&= \frac{10}{\varepsilon^2}(y_2-y_1) \label{eqn:y1}\\
	\frac{dy_2}{dt}&= \frac{1}{\varepsilon^2}(28y_1-y_2-y_1y_3)\\
	\frac{dy_3}{dt}&= \frac{1}{\varepsilon^2}(y_1y_2-\frac{8}{3}y_3)\label{eqn:y3}.
\end{align}
Here the slow variable $x$ is driven by a fast chaotic Lorenz system \citep{Lorenz63}. The slow dynamics (\ref{eqn:x}) describes an overdamped degree of freedom in a one-dimensional potential well $V(x) = x^4/4 - x^2/2$ which is being continually ``kicked'' by the fast chaotic motion of the Lorenz subsystem (\ref{eqn:y1})-(\ref{eqn:y3}). In the following we will set $\varepsilon^2=0.01$ unless specified otherwise.\\

\begin{figure}[htbp]
\centering
\includegraphics[width = \columnwidth]{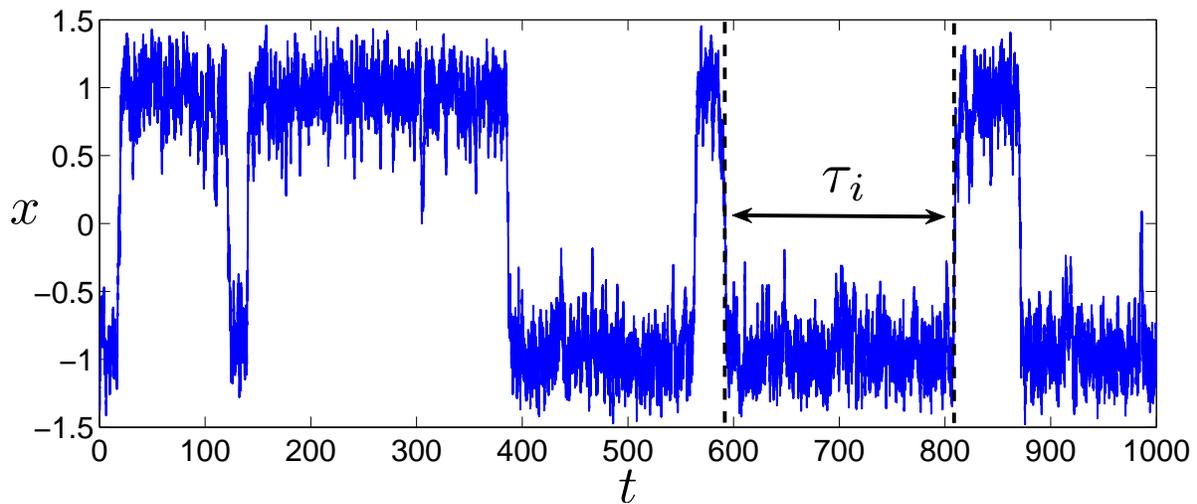}
\caption{Sample trajectory of the slow variable $x$ of the full deterministic 4D model (\ref{eqn:x})-(\ref{eqn:y3}). The dynamics appears to randomly switch between two metastable states with randomly distributed sojourn times $\tau_i$.}
\label{fig:taui}
\end{figure}

\noindent
In Figure~\ref{fig:taui} we show a trajectory of the slow variable $x$ obtained from a simulation of the full deterministic system (\ref{eqn:x})-(\ref{eqn:y3}). The trajectory of the slow variable appears to randomly switch between two metastable states around the minima of the potential $V(x)$ at $x^\star=\pm1$. In the following Section we will derive a reduced stochastic differential equation which describes the effective slow dynamics.

%


\subsection*{Derivation of the stochastic climate model}
The ergodicity of the fast chaotic Lorenz equation \citep{Tucker99} and its mixing property \citep{Luzzatto05} allows for an application of stochastic model reduction techniques, by which the fast chaotic degrees of freedom are parametrized by a one-dimensional stochastic process. This can be heuristically justified provided the fast processes decorrelate rapidly enough; then the slow variables experience the sum of uncorrelated fast dynamics during one slow time unit. According to the (weak) Central Limit Theorem this converges to approximate Gaussian noise in the limit when the time scale separation becomes infinite. We will formalize this and apply stochastic singular perturbation theory (homogenization) \citep{Khasminsky66,Kurtz73,Papanicolaou76,Givonetal04,PavliotisStuart} to deduce the following reduced stochastic 1D climate model for the slow $x$-dynamics
\begin{equation}
\frac{dX}{dt} = X (1- X^2) + \sigma \frac{dW}{dt} \label{eqn:climate}
\end{equation}
with 1-dimensional Wiener process $W$, where $\sigma$ is given by the total integral of the autocorrelation function of the fast $y_2$ variable  with
\begin{align}
\frac{\sigma^2}{2} = \left(\frac{4}{90}\right)^2 
\int_0^\infty 
\{ \lim_{T \rightarrow \infty} \frac{1}{T} \int_0^T y_2(s)y_2(t+s)ds \}\,dt\; .
\label{e.sigma2}
\end{align}

Rather than studying the system (\ref{eqn:x})-(\ref{eqn:y3}) directly, in stochastic homogenization one considers the associated Fokker-Planck or its adjoint backward Kolmogorov equation. Whereas the original ordinary differential equation is nonlinear, the latter ones are linear partial differential equations and can be treated with standard perturbation techniques, expanding in the small parameter $\varepsilon$ and studying solvability conditions of the linear equations at the respective orders of $\varepsilon$. The solvability conditions are given by Fredholm alternatives and can be evaluated using the ergodicty of the fast process.\\

We will analyze the system (\ref{eqn:x})-(\ref{eqn:y3}) in the framework of the backward Kolmogorov equation for the conditional expectation value of some sufficiently smooth observable $\phi(x,y)$ defined as
\[
v(x_0,y_0,t)=\mathbb{E}\left[\phi(x(t),y(t))\left|\right. x(0)=x_0,y(0)=y_0\right]\;
\]
with $y=(y_1,y_2,y_3)$. Here the expectation value is taken with respect to the ergodic measure induced by the fast dynamics of the chaotic Lorenz system. We drop the 0-subscripts for ease of exposition from here on. We study the following Cauchy problem for $t\in[0,\infty)$
\begin{align}
\frac{\partial }{\partial t}v(x,y,t) &= \L v(x,y,t)
\nonumber
\\
v(x,y,0) &= \phi(x,y)\; ,
\label{e.veq}
\end{align}
with the generator
\[
\L=\frac{1}{\varepsilon^2}\L_0+\frac{1}{\varepsilon}\L_1+\L_2\; ,
\]
where
\begin{align}
\L_0 &\equiv g(y) \cdot \nabla_y\\
\L_1 & \equiv \frac{4}{90}y_2 \frac{\partial}{\partial x}\\
\L_2  &\equiv x(1 - x^2) \frac{\partial}{\partial x}\; ,
\label{e.gen}
\end{align}
and $g(y)$ denotes the scaled vectorfield of the fast Lorenz system $g(y) = (10(y_2-y_1), (28y_1-y_2-y_1y_3), (y_1y_2-8y_3/3))$. We remark that equally we could have used the framework of the Fokker-Planck equation.

Pioneered by \cite{Kurtz73} and \cite{Papanicolaou76} a perturbation expansion can be made according to
\begin{align}
v(x,y,t)=v_0+\varepsilon v_1+\varepsilon^2v_2 + \cdots\; .
\label{e.vexp}
\end{align}
A recent exposition of the theory of homogenization and their applications is provided in \citep{Givonetal04,PavliotisStuart}. Substituting the series (\ref{e.vexp}) into the backward Kolmogorov equation (\ref{e.veq}) we obtain at lowest order, $\O(1/\eps^2)$,
\begin{eqnarray}
\L_0v_0=0\; .
\label{e.L0v0}
\end{eqnarray}
The chaotic dynamics of the fast Lorenz system, associated with the generator $\L_0$, is ergodic \citep{Tucker99}. Hence the expectation value does not depend on initial conditions, and we obtain
\[
v_0=v_0(x,t)
\]
as the only solution of (\ref{e.L0v0}).\\ Ergodicity implies the existence of a unique invariant density induced by the fast chaotic dynamics given by the unique solution of the associated Fokker Planck-equation 
\[
\L_0^\star\rho = 0\; ,
\]
where $\L_0^\star$ is the formal $L_2$-adjoint of the generator $\L_0$. We label this solution as $\rho_\infty(y)$.

At the next order, $\O(1/\eps)$, we obtain
\begin{align}
\L_0v_1=-\L_1v_0\; .
\label{e.Lv1}
\end{align}
To assure boundedness of $v_1$ (and thereby of the asymptotic expansion (\ref{e.vexp})) a solvability condition has to be satisfied prescribed by the Fredholm alternative. Equation (\ref{e.Lv1}) is solvable only provided the right-hand-side is in the space orthogonal to the (one-dimensional) null space of the adjoint $\L_0^\star$, i.e. if
\[
\langle \L_1v_0 \rangle_{\rho_\infty} 
= 
-\frac{4}{90}\langle y_2\rangle_{\rho_\infty} \frac{\partial }{\partial x}v_0(x,t) = 0\; ,
\]
where we introduced the average of an observable $h(x,y)$ over the fast ergodic density as $ \langle h \rangle_{\rho_\infty} := \int h(x,y) \rho_\infty(y) \, \d y$. Note that the vanishing of the average of the fast perturbation in the slow equation with respect to the invariant measure induced by the Lorenz system implies that classical averaging would only produce trivial reduced dynamics $\dot x = {\cal{O}}(\eps)$. Since $\langle y_2 \rangle_{\rho_\infty}=0$, there exists a solution of (\ref{e.Lv1}), which we can formally write as
\begin{equation}
v_1(x,y,t) = -\L_0^{-1}\L_1v_0 + R(x)\; ,
\label{e.v1}
\end{equation}
where $R(x)$ lies in the kernel of $\L_0$. 

At the next order, $\O(1)$, we obtain 
\begin{align}
\L_0v_2=\frac{\partial}{\partial t}v_0 - \L_1v_1 - \L_2 v_0\; ,
\label{e.v2}
\end{align}
which yields the desired evolution equation for $v_0$ as the associated solvability condition
\begin{equation}
\frac{\partial}{\partial t}v_0 
= 
\langle \L_2v_0\rangle_{\rho_\infty} +
\langle \L_1 v_1(x,t)\rangle_{\rho_\infty}
\; .
\label{e.BKgen}
\end{equation}
The reduced slow backward Kolmogorov equation is then
\begin{align}
\frac{\partial}{\partial t}v_0 
= x(1-x^2) \frac{\partial }{\partial x} v_0
- \left( \frac{4}{90}\right)^2 \langle y_2 \L_0^{-1} y_2 \rangle_{\rho_\infty}\frac{\partial^2}{\partial x^2} v_0 
\; .
\label{e.BKgen2}
\end{align}
This can be simplified further. For every function $h$ with $\langle h \rangle_{\rho_\infty}=0$ we define
\[
H(x,y) = -\int_0^\infty e^{\L_0 t} h \, dt\; ,
\]
which, upon using that $\L_0$ corresponds to an ergodic process, implies that $\L_0 H = h$ \citep{PavliotisStuart}. Hence we can evaluate
\begin{align*}
\langle y_2 \L_0^{-1} y_2 \rangle_{\rho_\infty}
&= - \int_0^\infty \langle y_2 e^{\L_0t} y_2 \rangle_{\rho_\infty}\, dt\\
&= - \int_0^\infty 
\{ \lim_{T \rightarrow \infty} \frac{1}{T} \int_0^T y_2(s)y_2(t+s) ds \}\,dt\;,
\; 
\end{align*}
where the last identity follows from employing Birkhoff's ergodic theorem. The reduced backward Kolmogorov equation (\ref{e.BKgen2}) can then be written as
\begin{align}
\frac{\partial}{\partial t}v_0 
= x(1-x^2) \frac{\partial }{\partial x} v_0
+ \frac{\sigma^2}{2}\frac{\partial^2}{\partial x^2} v_0 
\; ,
\label{e.BKgen3}
\end{align}
with $\sigma^2$ given by (\ref{e.sigma2}). Note that $R(x)$ does not contribute to the dynamics. We can therefore choose $R(x)=0$ in order to assure that $\langle v\rangle_{\rho_\infty} = \langle v_0\rangle_{\rho_\infty} + {\mathcal{O}}(\eps^2)$.

The slow reduced Langevin equation associated with the reduced backward Kolmogorov equation (\ref{e.BKgen3}) is then given by our stochastic 1D climate model (\ref{eqn:climate}).\\

\noindent
The unique invariant density of the gradient system (\ref{eqn:climate}) is readily determined as the unique stationary solution $\hat{\rho}(x)$ of the associated Fokker-Planck equation
\begin{align}
\frac{\partial}{\partial t} \rho(x) = \frac{\partial}{\partial x}\left(\frac{dV}{dx}\,\rho\right) + \frac{\sigma^2}{2}\frac{\partial^2}{\partial {x^2}}\rho\; .
\label{e.FP}
\end{align}
We find 
\begin{align}
\hat{\rho}(x)=\frac{1}{Z}e^{-\frac{2}{\sigma^2}V(x)}
\label{e.rhohat}
\end{align}
with 
\[
Z=\int_{-\infty}^\infty e^{-\frac{2}{\sigma^2}V(x)}dx\; .
\]
Note that the unique invariant density of the full deterministic 4D system (\ref{eqn:x})-(\ref{eqn:y3}) is now approximated by $\rho(x,y)=\hat{\rho}(x)\rho_{\infty}(y) + {\mathcal{O}}(\eps)$. 


\section{Parameter estimation}
\label{sec:paraest}
The value of the diffusion coefficient $\sigma$ given by (\ref{e.sigma2}) cannot be determined analytically for the reduced stochastic 1D climate model as the unique invariant density $\rho_{\infty}(y)$ of the fast Lorenz subsystem  (\ref{eqn:y1})-(\ref{eqn:y3}) is not explicitly known. We therefore need to evaluate the diffusion coefficient $\sigma$ numerically from simulations of the full deterministic 4D model (\ref{eqn:x})-(\ref{eqn:y3}) under the assumption that its slow dynamics is well approximated by the stochastic 1D climate model (\ref{eqn:climate}). We do so by coarse-graining the full deterministic 4D system and estimating conditional averages \citep{Gardiner,Siegert98,Stemler07}. This method has been used in the meteorological community to study time series in diverse contexts such as synoptic variability of midlatitude sea surface winds, Icelandic meteorological data, planetary waves and paleoclimatic ice-core data \citep{Sura02,Sura03,EggerJonsson02,Berner05,Ditlevsen99}. This approach also allows us to determine the drift term of the stochastic 1D climate model, which we write for convenience here as
\begin{align}
dx = d(x)dt + \sigma dW\, ,
\label{e.langevin}
\end{align}
with drift term $d(x)=x(1-x^2)$. Numerically integrating the full deterministic 4D system (\ref{eqn:x})-(\ref{eqn:y3}) with a small time step $dt$, we create a long time series $x(t_i)$, $i=0,1,\cdots,n$ with $t_i=i\, dt$. To estimate the parameters of the Langevin system (\ref{e.langevin}) we subsample the time series $x(t_i)$ at a coarse sampling time $h\gg dt$. We choose $h$ such that during $h$ the finely sampled trajectory $x(t_i)$ performs roughly $3$-$4$ fast (smooth) undulations induced by the fast chaotic dynamics.\\ We estimate the diffusion coefficient $\sigma$ (and the drift term $d(x)$) by partitioning the state space into bins $X_i$ of bin size $\Delta X$. We coarse grain by mapping $x$ into $X$ for $x \in (X,X+\Delta X)$. To estimate the diffusion coefficient we define the average conditioned on the bins $[X,X+\Delta X]$
\begin{align}
S(X) &= 
\frac{1}{h}
\langle
(x^{n+1}-x^{n})^2
\rangle
\Big|_{x^n \in (X,X+\Delta X)}
\; ,
\end{align}
where $x^n=x(nh)$. The angular brackets denote a conditional ensemble average over microscopic realizations $x(t_i)$ which fall into a coarse macroscopic bin $[X,X+\Delta X]$. In the limit when the deterministic 4D dynamics can be approximated by the stochastic 1D climate model (\ref{e.langevin}) and for small values of the sampling time $h$, we can evaluate
\[
S(X) = \sigma^2 + X^2(1-X^2)^2h \; ,
\]
where we used $\langle dW \rangle=0$. In the limit $h\to 0$ we estimate $S(X)=\sigma^2$. Similarly the drift $d(x)$ can be estimated by
\begin{eqnarray}
D(X) = 
\frac{1}{h}
\langle
(x^{n+1}-x^{n})
\rangle
\Big|_{x^n \in (X,X+\Delta X)}
\; ,
\end{eqnarray}
with $D(X)\approx d(x)$ in the limit $h\to 0$.\\

\noindent
There are several pitfalls one may encounter using this method to estimate the effective drift and diffusion coefficients of  chaotic deterministic differential equations related to the choice of the sampling time $h$. (Sampling issues for purely stochastic multiscale systems were already reported in \cite{Pavliotis07}.) If $h$ is chosen smaller than the typical time scales on which the fast subsystem decorrelates, the diffusion coefficient does not exist and $S(X)$ will be close to zero in the limit $h\to 0$. On the contrary, if $h$ is chosen too large the estimator for the diffusion coefficient will be swamped by the deterministic part of the dynamics and we will obtain $S(X)\approx X^2(1-X^2)^2h$. Similarly there are problems with estimating the drift $D(X)$ for sampling times $h$ taken too large. For such subsampling times $h$, the coarse-grained dynamics will randomly switch from bin to bin according to the invariant density $\hat{\rho}(x)$ given by (\ref{e.rhohat}). Since in our case $\hat{\rho}(x)$ is symmetric, we can write the continuous approximation of the estimator for the drift as
\[
D(X)=\frac{1}{h}\int(Y-X)\, \hat{\rho}(Y)\, dY = -\frac{X}{h}\; .
\]
This may lead to erroneous classifications of stochastic processes as Ornstein-Uhlenbeck processes. We remark that if the time series is not sufficiently long, the estimated drift coefficient in our case of a bimodal probability density function will be erroneously estimated as $D(X)=-(X-X_0)/h$ due to the presence of the two metastable states around $X_0=\pm1$.\\

\noindent
This procedure can be readily extended to study multi-dimensional stochastic processes where the bins would be hypercubes; the interpretation of the results in higher dimensions and the deducing functional dependencies may however be difficult. We remark that one may also use Kalman filters to estimate the parameters \citep{Cornick09}, or estimate the generator directly \citep{Crommelin06}.\\

\begin{figure}[htbp]
\centering
\includegraphics[width = 0.55\columnwidth]{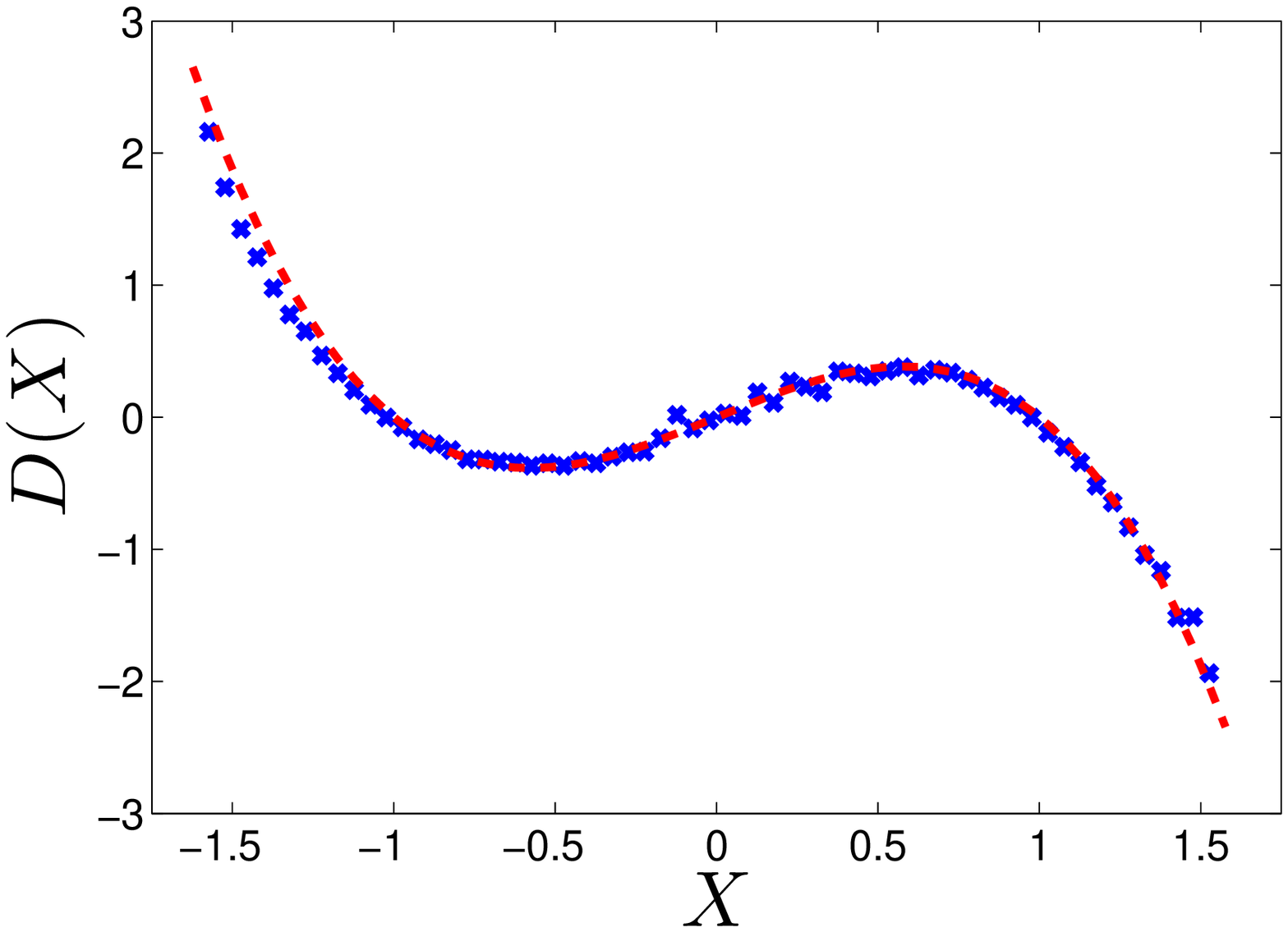}\\
\includegraphics[width = 0.55\columnwidth]{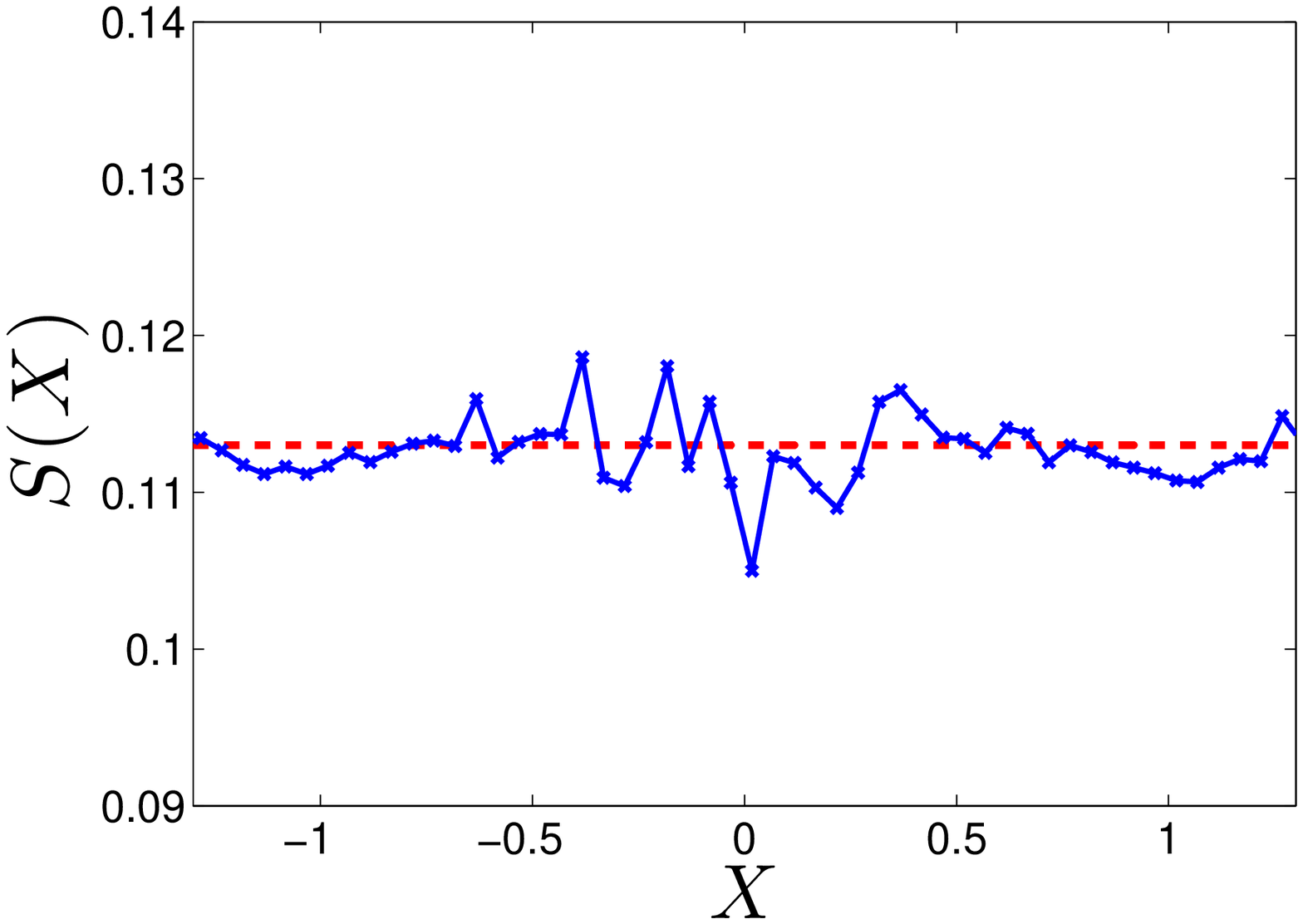}\\
\includegraphics[width = 0.55\columnwidth]{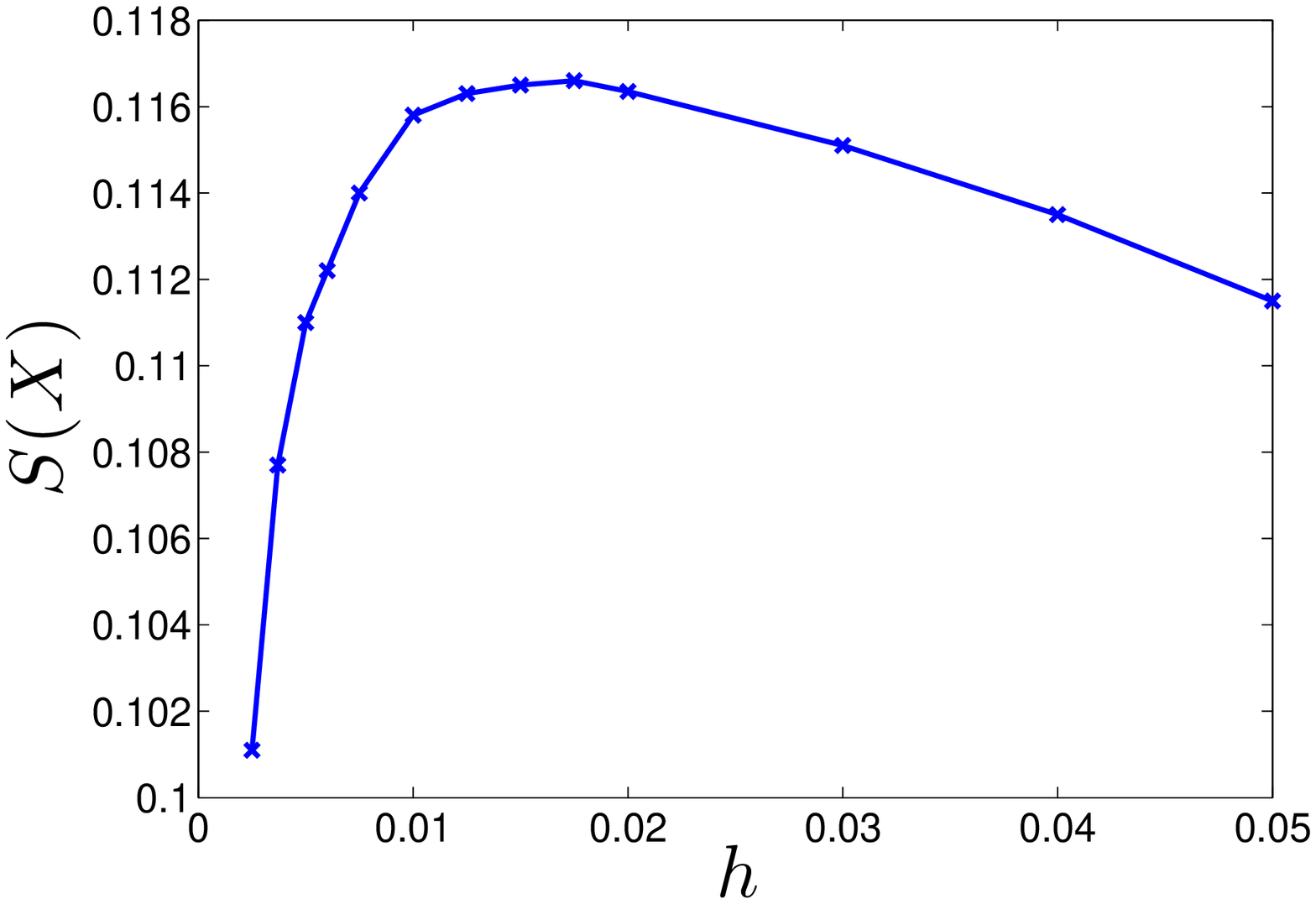}
\caption{Parameter estimation for the (a) drift term $D(X)$, and (b) diffusion coefficient $S(X)$ from a long trajectory of the full deterministic 4D system (\ref{eqn:x})-(\ref{eqn:y3}) with $\epsilon^2=0.0005$ and sample interval $h=0.005$. The dashed 
line depicts the theoretically expected drift from the stochastic 1D climate model (\ref{eqn:climate}) in (a), and the constant diffusion coefficient $\sigma^2=0.113$ in (b). (c): Diffusion coefficient $S(X)$ as a function of the sampling time $h$.}
\label{f.pest}
\end{figure}

\noindent
In Figure~\ref{f.pest} we show results from the parameter estimation procedure. We choose a coarse-grain bin size of $\Delta X=0.05$ and sample the slow variable $x$ at constant sampling intervals $h=0.005$ corresponding to $4$ fast oscillations. We used a trajectory of the full deterministic 4D system (\ref{eqn:x})-(\ref{eqn:y3}) with $\varepsilon^2=0.0005$ and evolved until $T=3\times10^4$ time units. We confirm the 1D climate model (\ref{eqn:climate}) by estimating the drift coefficient clearly as 
\[
D(X) = X(1-X^2)\; ,
\]
and estimating a constant diffusion coefficient $S(X)=\sigma^2=0.113$. We note that this is slightly smaller than the value $\sigma^2=0.126$ which was found by \cite{Givonetal04} using a different method.\\ In Figure~\ref{f.pest}(c) we show how the diffusion coefficient $S(X)$ depends on the sampling time $h$ as discussed above. It is clearly seen that the diffusion coefficient $\sigma$ cannot be unambigously determined (at least not using this method). We will investigate in Section~\ref{sec:sensitivity} how sensitive the statistics of the 1D climate model (\ref{eqn:climate}) is to this degree of uncertainty in the estimation of the diffusion coefficient.

%


\section{Time scales}
\label{sec:timescales}
An important parameter in data assimilation is the observation interval $\Delta \tobs$. We will see in Section~\ref{sec:results} that skill improvement of the analysis is dependent on $\Delta \tobs$. We therefore now analyze several characteristic time scales of the deterministic 4D toy model (\ref{eqn:x})-(\ref{eqn:y3}). 
In particular we will focus on the slow variable $x$.

\subsection{Autocorrelation time} 
We first estimate the decorrelation time $\tau_{\rm corr}$, which is linked to the decay rate of the autocorrelation function of the slow variable
\begin{align}
C(\tau) = \lim_{T\to\infty}\frac{1}{T}\int_0^T x(s)x(\tau+s) \, ds\; .
\label{e.ACF}
\end{align}
The autocorrelation time is estimated as the $e$-folding time of $C(\tau)$. The autocorrelation function $C(\tau)$ can be estimated from a long trajectory of $x$ obtained from a simulation of the full deterministic 4D model (\ref{eqn:x})-(\ref{eqn:y3}). For small values of $\tau$ it decays approximately exponentially with a measured decay rate of $0.00481$ corresponding to an $e$-folding time of $\tau_{\rm corr}=208$ time units.\\



\subsection{Sojourn and exit times}
\begin{figure}[htbp]
\centering
\includegraphics[width = 0.7\columnwidth]{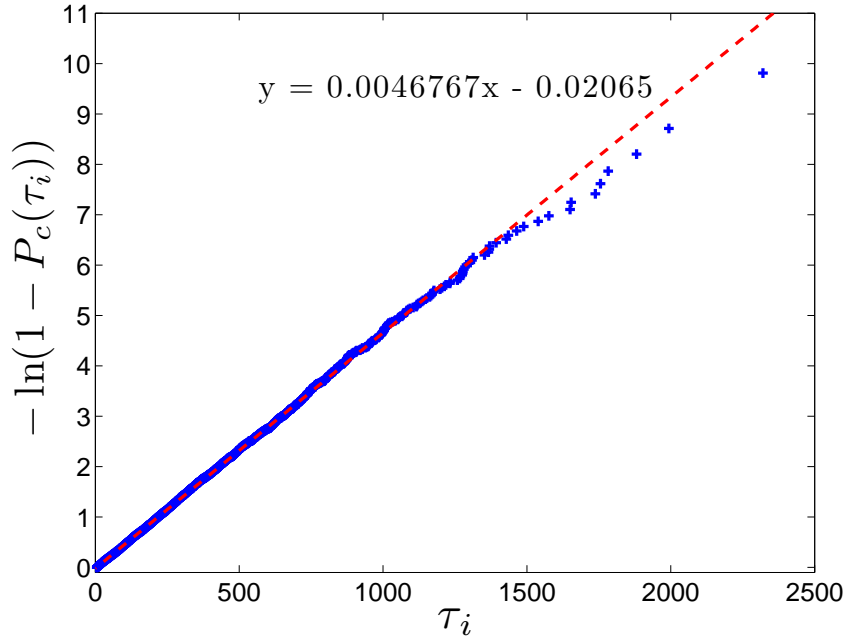}
\caption{Log-plot of the normalised histogram of sojourn times $\tau_i$ of the slow variable $x$ in each metastable state around $x^\star=\pm 1$ for the full deterministic 4D system (\ref{eqn:x})-(\ref{eqn:y3}). The dashed line shows a least-square fit. Using (\ref{e.poisson_cdf}) the mean sojourn time can be estimated via the inverse of the slope of the least-square fit as $\bar{\tau} \approx 213.8$.}
\label{fig:sojourn}
\end{figure}

As a second time scale we consider the mean sojourn time 
\begin{align}
\bar{\tau} = \frac{1}{M}\sum_{i=1}^M \tau_i\; ,
\label{e.msav}
\end{align}
where the sojourn times $\tau_i$ measure the times spent in one of the slow metastable states around $x^\star=\pm1$ as depicted in Figure~\ref{fig:taui}. We expect this time scale to strongly correlate with the autocorrelation time scale $\tau_{\rm corr}$ as it is random transitions between the slow regimes which cause decorrelation of the slow variable. We numerically estimate the mean sojourn time using two separate methods. Firstly, we estimate the mean sojourn time directly from a long trajectory of the slow variables $x$ using (\ref{e.msav}) as $\bar\tau\approx 218.2$ time units. Secondly, the succession of rapid transitions and relatively long residence times in the potential wells suggests that successive jumps between metastable states can be treated as independent random events, and that the sojourn times are a Poisson process with
cumulative probability distribution function
\begin{equation}
P_c(\tau_i) = 1 - \exp\left( -\frac{\tau_i}{\bar{\tau}} \right)\; .
\label{e.poisson_cdf}
\end{equation}
In Figure \ref{fig:sojourn} we show the empirical ranked histogram $P_c(\tau_i)$ for the sojourn times measured from a long trajectory $x(t)$ of the full deterministic 4D system (\ref{eqn:x})-(\ref{eqn:y3}), allowing us to determine the mean sojourn time $\bar{\tau} \approx 213.8$ time units. Calculating the mean sojourn time by either the average of individual sojourn times (\ref{e.msav}) or via the Poisson process approximation (\ref{e.poisson_cdf}) yields results differing by only $2\%$, indicating that for $\eps^2=0.01$ the fast chaotic Lorenz subsystem has almost fully decorrelated and that the rare transitions between the metastable states are approximately a Poisson process. In the following we will use the average of the two obtained values of the mean sojourn time and set $\bar \tau =216.0$.

\noindent
These numerically obtained results for the full deterministic 4D system (\ref{eqn:x})-(\ref{eqn:y3}) compare well with the value calculated analytically for the stochastic 1D climate model (\ref{eqn:climate}) using Kramer's theory \citep{Kramers40} to calculate the first exit time
\[
\tau_e = \frac{1}{2} \bar{\tau}\; .
\]
We define the first exit time $\tau_e$ to be the average time it takes from the potential well at $x=\pm1$ to reach the saddle of the potential $V(x)$ at $x=0$. To calculate the exit time $\tau_e$ we solve the following Cauchy problem (see for example \cite{Zwanzig})
\begin{align}
\L_{\rm clim}\tau_e = -1\; ,
\label{e.FET}
\end{align}
where 
\begin{align*}
\L_{\rm clim}= -V^\prime \partial_x + \frac{\sigma^2}{2}\partial_{xx}
\end{align*}
is the generator of the stochastic 1D climate model (\ref{eqn:climate}), with the boundary conditions $\partial_x \tau_e(x=\pm1)=0$ and $\tau_e(0)=0$. Note that the first exit times for the symmetric potential $V$ out of the potential wells at $x=-1$ or $x=1$ are identical. Upon using the boundary conditions, we solve (\ref{e.FET}) to obtain
\[
\tau_e = \beta\int_{-1}^0dy\, e^{\beta V(y)} \int_{-1}^ye^{-\beta V(z)} dz\; ,
\]
with $\beta=2/\sigma^2$, which can be numerically evaluated to yield $\tau_e=117.8$ for $\sigma^2 = 0.113$.
The exact analytical value calculated from the reduced stochastic 1D climate model compares well with the numerically estimated values of the first exit times of the full deterministic 4D model $\tau_e=108.0$, confirming the accuracy of the homogenized model to faithfully reproduce the statistics of the full parent model.\\


\subsection{Transit time}
As a third characteristic time scale, we study the mean transit time $\tau_t$ which is the average over the shortest transit time $\tau_{t,i}$ between the two slow metastable states around $x^* = \pm 1$. We define this time to be the mean of 
\begin{align}
\tau_{t,i} = \min \{ (t_i - t_j) | x(t_i) = \pm 1 \land  x(t_j) = \mp 1\}\, .
\label{e.tauti}
\end{align}
For a long trajectory containing 5000 transitions between the slow metastable states we numerically estimate $\tau_t = 5.90$ for the full deterministic 4D model (\ref{eqn:x})-(\ref{eqn:y3}). We found that the transit time does not vary greatly with the value of $\eps$.

\noindent
As for the mean sojourn time $\bar{\tau}$, the transit time $\tau_t$ can be analytically calculated for the stochastic 1D climate model (see for example \cite{Gardiner}) as
\begin{align}
\tau_t = 
4
\frac{\Pi_{+1}(0)-\Pi_{-1}(0)}{\sigma^2 \hat \rho_{-1}(0)}\; ,
\label{e.taut}
\end{align}
where 
\begin{align*}
\Pi_{+1}(x)&=\hat \rho_{+1}(x)\int_x^{1}dx^\prime \hat\rho^{-1}(x^\prime)\int_{-1}^{x^\prime} \hat \rho_{-1}(z)\hat \rho(z) dz\\
\Pi_{-1}(x)&=\hat \rho_{-1}(x)\int_{-1}^xdx^\prime \hat\rho^{-1}(x^\prime)\int_{-1}^{x^\prime} \hat \rho_{-1}(z)\hat \rho(z) dz
\; ,
\end{align*}
with the splitting probabilities
\begin{align*}
\rho_{-1}(x)&=\frac{\int_x^1dz\hat \rho^{-1}(z)}{\int_{-1}^1dz\hat \rho^{-1}(z)}\\
\rho_{+1}(x)&=\frac{\int_{-1}^xdz\hat \rho^{-1}(z)}{\int_{-1}^1dz\hat \rho^{-1}(z)} \; .
\end{align*}
For the stochastic 1D climate model with $\sigma^2=0.113$ we evaluate this numerically to be $\tau_t=5.66$ which again compares well with the numerically obtained value of the transit time for the full deterministic 4D system. 




\section{Validity of the stochastic climate model}
\label{sec:sensitivity}

\noindent
In this Section we numerically investigate to what degree the stochastic 1D climate model (\ref{eqn:climate}) faithfully reproduces the slow dynamics of the full deterministic 4D system (\ref{eqn:x})-(\ref{eqn:y3}). It is clear that such a correspondence can only be of a statistical nature.  The close correspondence of the time scales of the full deterministic 4D system and of the stochastic 1D climate model for $\sigma^2=0.113$ and $\eps^2=0.01$ already indicates the validity of the stochastic 1D climate model in reproducing statistical aspects of the slow dynamics. The rigorous theory by \cite{Kurtz73} and \cite{Papanicolaou76} assures weak convergence of the stochastic climate model in the limit of $\eps\to 0$. Here we discuss the effect of finite time scale separation $\eps$. Furthermore we study the sensitivity of the climate model to uncertainties in the diffusion coefficient $\sigma^2$ as we have encountered in Section~\ref{sec:paraest}. In the Appendix we will also investigate the sensitivity to possible uncertainties in the drift term.\\ 


\subsection{Probability density function}

\begin{figure}[htbp]
\centering
\includegraphics[width = 0.7\columnwidth]{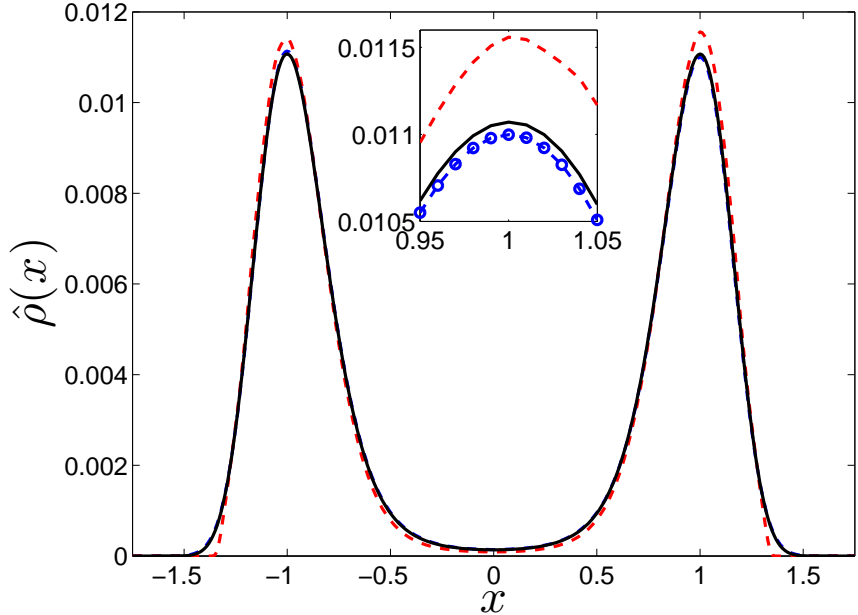}
\caption{Normalised empirical densities for the slow variable $x$ of the full deterministic 4D model (\ref{eqn:x})-(\ref{eqn:y3}) with $\eps^2 = 0.1$ (dashed line), $\eps^2 = 0.01$ (solid line) and $\eps^2 = 0.0005$ (circles). The empirical densities were obtained from long trajectories integrated to time $T = 10^9$ time units.}
\label{f.pdf_eps}
\end{figure}

\noindent
We first examine how the probability density function varies for different values of the time scale separation parameter $\eps$. In Figure \ref{f.pdf_eps} we show empirical densities for the slow variable $x$ obtained from a long simulation of the full deterministic 4D system (\ref{eqn:x})-(\ref{eqn:y3}) for different values of the time scale separation $\eps$. The obtained estimates of the empirical invariant densities exhibit a bimodal structure reflecting the two slow metastable regimes near $x^\star=\pm 1$.  The probability density functions are reasonably insensitive to changes in $\eps^2$, with the maxima of the densities at $x^\star=\pm1$ differing for $\eps^2 = 0.0005$ and $\eps^2 = 0.01$ by less than $1\%$, and by approximately $5\%$ for $\eps^2 = 0.01$ and $\eps^2 = 0.1$. This numerically demonstrates the weak convergence of solutions of the homogenized stochastic 1D climate model to solutions of the full deterministic slow dynamics.\\

\begin{figure}[htbp]
\centering
\includegraphics[width = 0.7\columnwidth]{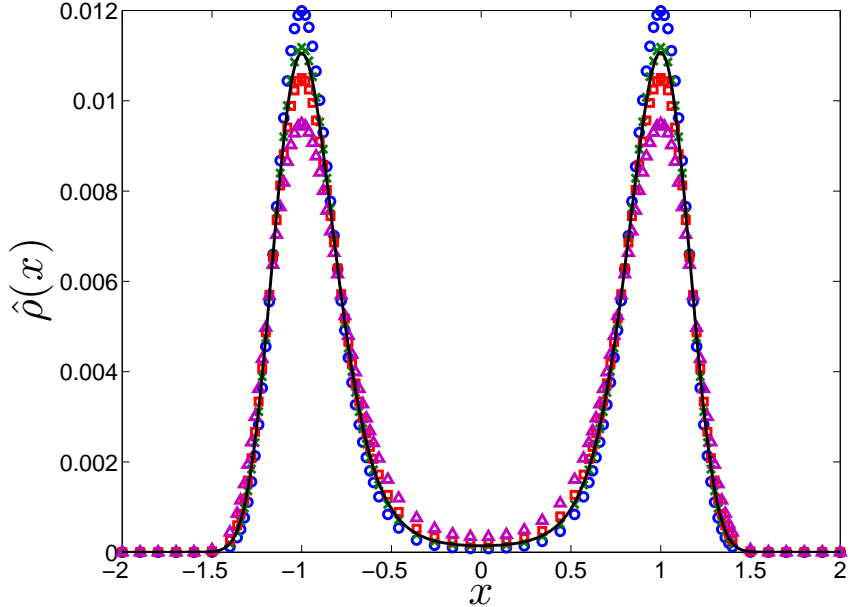}
\caption{Invariant density (\ref{e.rhohat}) for the stochastic 1D climate model (\ref{eqn:climate}), with $\sigma^2 = 0.1$ (circles), $\sigma^2 = 0.113$ (crosses), $\sigma^2 = 0.126$ (squares) and $\sigma^2 = 0.15$ (triangles), and empirical density for the slow variable $x$ of the full deterministic 4D model (\ref{eqn:x})-(\ref{eqn:y3}) with $\eps^2 = 0.01$ (solid line).}
\label{f.pdf_sigma}
\end{figure}

\noindent
Next we show in Figure~\ref{f.pdf_sigma} how sensitive the empirical density of the stochastic 1D climate model (\ref{eqn:climate}) is to changes in the diffusion coefficient $\sigma^2$. Unlike for changes in the time scale parameter $\eps$, the 1D climate model is quite sensitive to changes in the diffusion coefficient. The invariant density of the full deterministic 4D model is best estimated by the stochastic 1D climate model with $\sigma^2 = 0.113$.\\

%
%


\subsection{Time scales}

\noindent
Differences in the probability density function imply different statistics of the dynamics. In particular, we will now examine how the characteristic time scales of the full deterministic 4D model (\ref{eqn:x})-(\ref{eqn:y3}) change with the time scale separation $\eps$, and how they change for the stochastic 1D climate model (\ref{eqn:climate}) with varying diffusion coefficients $\sigma^2$.\\ 

\noindent
Consistent with our earlier observation that the probability density function does not vary greatly with the time scale separation parameter $\eps$ (provided it is sufficiently small), we confirm here that the time scales are insensitive to a decrease in $\eps$ for $\eps^2<0.01$. In particular the transit time $\tau_t$ exhibits very little variation with $\eps$.\\

%
%

\begin{figure}[htbp]
\centering
\includegraphics[width = 0.7\columnwidth]{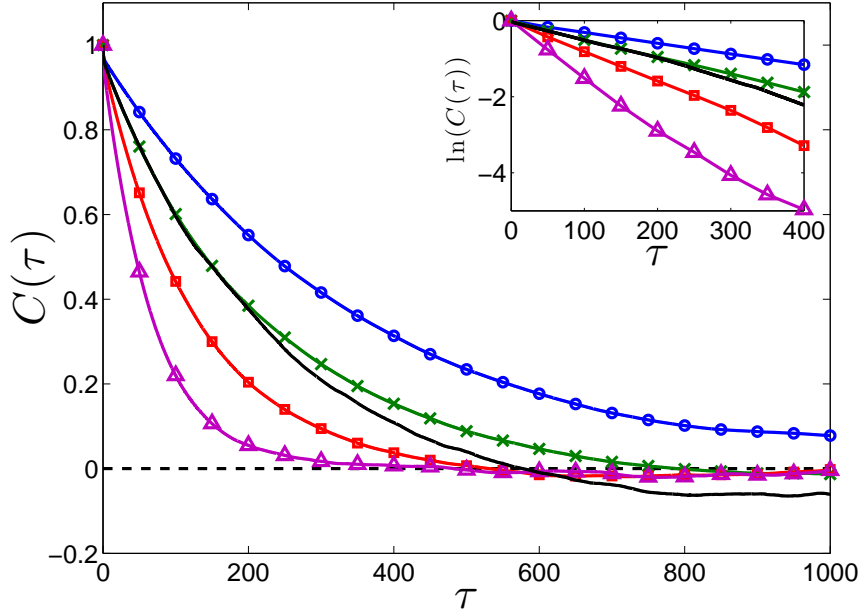}
\caption{Autocorrelation function $C(\tau)$ as a function of the time lag $\tau$ for the full deterministic 4D model (\ref{eqn:x})-(\ref{eqn:y3}) (solid line) and for the stochastic 1D climate model (\ref{eqn:climate}) with $\sigma^2 = 0.1$ (circles), $\sigma^2 = 0.113$ (crosses), $\sigma^2 = 0.126$ (squares) and $\sigma^2 = 0.15$ (triangles). 
}
\label{fig:ACF}
\end{figure}

\noindent
We find that with the exception of the transit time $\tau_t$, the characteristic time scales of Section \ref{sec:timescales} are very sensitive to uncertainties in the diffusion coefficient $\sigma^2$. For example, in Figure \ref{fig:ACF} the autocorrelation function $C(\tau)$ is shown as estimated from a long trajectory of $x$ obtained from a simulation of the full deterministic 4D model (\ref{eqn:x})-(\ref{eqn:y3}), and from simulations of the stochastic 1D climate model (\ref{eqn:climate}) with different values of the diffusion coefficient $\sigma^2$. Using $\sigma^2 = 0.113$ in (\ref{eqn:climate}) produces the best fit to the shape of $C(\tau)$ for the full deterministic 4D model (\ref{eqn:x})-(\ref{eqn:y3}).\\

%

\noindent
Table \ref{tbl:timescales_diffusion} lists the values of the characteristic time scales for the stochastic 1D climate model (\ref{eqn:climate}) for various values of the diffusion coefficient $\sigma^2$. We show the percentage difference of each value with that for the full deterministic 4D model (\ref{eqn:x})-(\ref{eqn:y3}) with $\eps^2 = 0.01$ in brackets. All time scales vary by approximately a factor of 5 over the range of values indicated except for the transit time which varies by just $13\%$ and approximates the transit time of the full deterministic model better as $\sigma^2$ decreases. The transit time is determined by the rate of divergence associated with the unstable fixpoint at $x=0$ of the deterministic drift term, making it relatively insensitive to changes in the diffusion coefficient. The transit times of the full deterministic model however are slightly overestimated, since the definition (\ref{e.tauti}) does not take into account the skewness of the probability density function (see Figure~\ref{f.pdf_eps}) which implies that the trajectory spends more time within the range $x\in[-1,1]$ than outside this range. The table indicates that all time scales (except the transit time $\tau_t$) are best approximated by the stochastic 1D climate model with $\sigma^2 = 0.113$, consistent with the better approximation of the probability density function for that value of $\sigma^2$. However, we will see in Section \ref{sec:results} that in the context of ensemble data assimilation it is not necessarily ideal to use this value for $\sigma^2$; in fact it is preferable to use models with larger diffusion to create a more reliable ensemble with superior analysis skill.

\begin{table*}[htbp]
\caption{Characteristic time scales of the stochastic 1D climate model (\ref{eqn:climate}) for different values of the diffusion coefficient $\sigma^2$. In brackets we indicate the error when compared to values obtained from the full deterministic 4D model (\ref{eqn:x})-(\ref{eqn:y3}) with $\eps^2 = 0.01$.}
\label{tbl:timescales_diffusion}
\begin{center}
\begin{tabular}{|c|c|c|c|c|}
\hline
  &  $\sigma^2=0.1$ & $\sigma^2=0.113$ & $\sigma^2=0.126$ & $\sigma^2=0.15$\\
\hline
$\tau_{\rm corr}$ & 353.9 (70.1\%)  & 221.7 (6.6\%) & 129.0 (61.2\%) & 70.5 (195\%) \\
$\tau_e$ &   205.7 (90.5\%) & 117.8 (9.1\%) & 75.6 (42.6\%) & 40.8 (164.7\%)\\
$\tau_t$&  5.86 (0.07\%) & 5.66 (4.2\%) & 5.48 (7.7\%) & 5.17 (14.1\%)\\
\hline
\end{tabular}
\end{center}
\end{table*}
%


\section{Ensemble Kalman filtering}
\label{sec:EnKF}
We briefly introduce how data assimilation is performed in an ensemble filter framework. Given an $N$-dimensional dynamical system
\begin{align}
\dot{\z}_t = {\bf{F}}(\z_t)\; ,
\label{e.ODE}
\end{align}
which is observed at discrete times $t_i=i \Delta \tobs$, data assimilation aims at producing the best estimate of the current state given a typically chaotic, possibly inaccurate model $\dot{\z}=f(\z)$ and noisy observations of the true state $\z_t$ \citep{Kalnay}.
 
Observations ${\xobs}\in \mathbb{R}^n$ are expressed as a perturbed truth according to 
\[
{\xobs}(t_i) = \H \z_t(t_i) + {\robs}\, ,
\]
where the observation operator $\H:\mathbb{R}^N\to \mathbb{R}^n$ maps from the whole space into observation space, and $\robs \in \mathbb{R}^n$ is assumed to be i.i.d. observational Gaussian noise with associated error covariance matrix $\Robs$.

In an ensemble Kalman filter (EnKF) \citep{Evensen94,Evensen} an ensemble with $k$ members $\z_k$
\[
\Z=\left[ \z_1,\z_2,\dots,\z_k \right]
\in \mathbb{R}^{N \times k}
\]
is propagated by the dynamics (\ref{e.ODE}) according to
\begin{equation}
{\dot{\Z}} = {\bf{f}}(\Z)\; ,
\quad
{\bf{f}}(\Z) =\left[ f(\z_1),f(\z_2),\dots,f(\z_k) \right]
\in \mathbb{R}^{N\times k} \; . \label{eqn:f}
\end{equation}
This forecast ensemble is split into its mean $\bar{\z}_f$ and ensemble deviation matrix $\Z'_f$. The ensemble deviation matrix $\Z'_f$ can be used to approximate the ensemble forecast error covariance matrix via
\begin{equation}
\P_f(t)
= 
\frac{1}{k-1}\Z^\prime(t)\left[\Z^\prime(t)\right]^T
\in \mathbb{R}^{N\times N}\; .
\label{eqn:Pf}
\end{equation}
Note that $\P_f(t)$ is rank-deficient for $k<N$ which is the typical situation in numerical weather prediction where $N$ is of the order of $10^9$ and $k$ of the order of $100$.

Given the forecast mean $\bar{\z}_f$, the forecast error covariance $\P_f$ and an observation $\zobs$, an analysis mean is produced by minimizing the cost function
\[
J({\z}) = 
 \frac{1}{2}({\z}-\bar{\z}_f)^T\PfI(\z-\bar{\z}_f)
+ \frac{1}{2}(\zobs-\H\z)^T\RobsI(\zobs-\H\z) 
\]
which penalizes distance from both the forecast mean and the observations with weights given by the inverse of the forecast error covariance and the observational noise covariance, respectively. The analysis mean is readily calculated as the critical point of this cost function with
\begin{equation}
\label{eqn:zaens}
{\bar{\z}}_a 
= 
{\bar{\z}}_f 
+ \KR\left[{\zobs}  - \H{\bar{\z}}_f\right] 
\end{equation}
where
\begin{align}
\KR &= \P_f \H^T\left( \H \P_f\H^T + \Robs\right)^{-1} \nonumber \\
\P_a &= \left(\Id - \KR\H\right) \P_f\; .
\end{align}
Using the Shermann-Morrison-Woodbury formula \citep{Golub} we may rewrite the Kalman gain matrix and the analysis covariance matrix in the more familiar -- though computationally more complex -- form as $\KR = \P_a\H^T\RobsI$ and 
\begin{align}
\label{e.kirchoff}
\P_a = \left(\PfI + \H^T\RobsI \H\right)^{-1}\, .
\end{align}
To determine an ensemble $\Z_a$ which is consistent with the analysis error covariance $\P_a$, and satisfies
\[
\P_a
= 
\frac{1}{k-1}\Z_a^\prime\left[\Z_a^\prime\right]^T\; ,
\]
where the prime denotes the deviations from the analysis mean, we use the method of ensemble square root filters \citep{Simon}. In particular we use the method proposed in \citep{Tippett03,Wang04}, the so called ensemble transform Kalman filter (ETKF), which seeks a transformation $\S \in \mathbb{R}^{k\times k}$ such that the analysis deviation ensemble $\Z_a^\prime$ is given as a deterministic perturbation of the forecast ensemble $\Z_f$ via 
\begin{equation}
\Z_a^\prime=\Z_f^\prime \S \; .
\label{e.S}
\end{equation}
Alternatively one could choose the ensemble adjustment filter \citep{Anderson01} in which the ensemble deviation matrix $\Z_f^\prime$ is pre-multiplied with an appropriately determined matrix $\A\in \mathbb{R}^{N\times N}$. 
Note that the matrix $\S$ is not uniquely determined for $k<N$. The transformation matrix $\S$ can be obtained by \citep{Wang04} 
\[
\S = 
{\bar{\C}}
\left(\Id_k + {\bar{\Gam}} \right)^{-\frac{1}{2}} 
{\bar{\C}}^T\;.
\]
Here $\C\Gam\C^T$ is the singular value decomposition of
\[
{\bf{U}} 
= \frac{1}{k-1}
{\Z_f^\prime}^T
\H^T\RobsI\H
\Z_f^\prime\; .
\]
The matrix ${\bar{\C}}\in \mathbb{R}^{k\times (k-1)}$ is obtained by erasing the last zero column from $\C\in\mathbb{R}^{k\times k}$, and ${\bar{\Gam}} \in\mathbb{R}^{(k-1)\times (k-1)}$ is the upper left $(k-1)\times(k-1)$ block of the diagonal matrix $\Gam\in\mathbb{R}^{k\times k}$. The deletion of the $0$ eigenvalue and the associated columns in $\C$ assure that $\Z_a^\prime = \Z_a^\prime \S$ and therefore that the analysis mean is given by ${\bar{\z}}_a$. This method assures that the mean is preserved under the transformation \citep{Wang04} which is not necessarily true for general square root filters. We note that the continuous Kalman-Bucy filter can be used to calculate $\Z_a^\prime$ without using any computations of matrix inverses, which may be advantageous in high-dimensional systems \citep{BGR09}.\\

\noindent
A new forecast is then obtained by propagating $\Z_a$ with the nonlinear forecast model to the next time of observation, where a new analysis cycle will be started.\\

\noindent
In the next Section we perform data assimilation for the toy model introduced in Section~\ref{sec:model}. We examine when and why the stochastic 1D climate model (\ref{eqn:climate}) can be used as a forecast model to improve the analysis skill. In particular, we will relate the range of validity of the 1D climate model to the relation between the observation interval and the characteristic time scales introduced in Section~\ref{sec:timescales}.\\


\section{Stochastic climate model as forecast model in data assimilation}
\label{sec:results}

\noindent
We investigate how the ETKF as described in the previous Section performs when either the full 4D deterministic system (\ref{eqn:x})-(\ref{eqn:y3}) or the reduced stochastic 1D climate model (\ref{eqn:climate}) is used as a forecast model, when the truth evolves according to the full deterministic 4D model. We are concerned here with the following questions: Can the dimension reduced stochastic climate model produce a better analysis than the full deterministic model? And if so, under what circumstances and why?\\  We generate a truth $\z_t=(x_t,{y_1}_t,{y_2}_t,{y_3}_t)$ by integrating (\ref{eqn:x})-(\ref{eqn:y3}) using a fourth order Runge-Kutta scheme with timestep $dt = \eps^2/20$. Observations of the slow variables only are then generated at equidistant times separated by the observation interval $\Delta \tobs$ according to $x_{\rm obs}= x_t + \eta$ where $\eta\sim{\cal{N}}(0,\sqrt{R_{\rm obs}})$ with $R_{\rm obs}=0.5\, \sigma^2$ and $\sigma^2=0.126$. Note that this value is slightly larger than the optimal value $\sigma^2 = 0.113$ which was found in the previous section to best approximate the full deterministic 4D model. \\ To integrate the forecast model we use a fourth order Runge-Kutta scheme with timestep $dt = \eps^2/20$ for the full deterministic 4D model with $\eps^2=0.01$, and an Euler-Maruyama scheme for the reduced stochastic 1D climate model using the same timestep. Unless stated otherwise we use $\sigma^2=0.126$ for the stochastic 1D climate model. We include a spin-up time of 100 analysis cycles. We perform twin experiments, where both forecast models are started from the same initial conditions, and are then subsequently assimilated using the same truth and observations over 5000 time units. In order to avoid rare occasions of filter divergence due to an underestimation of the forecast covariances we employ a 2\% variance inflation \citep{AndersonAnderson99}.\\ 

\begin{figure}[htbp]
\begin{center}
\includegraphics[width=\columnwidth]{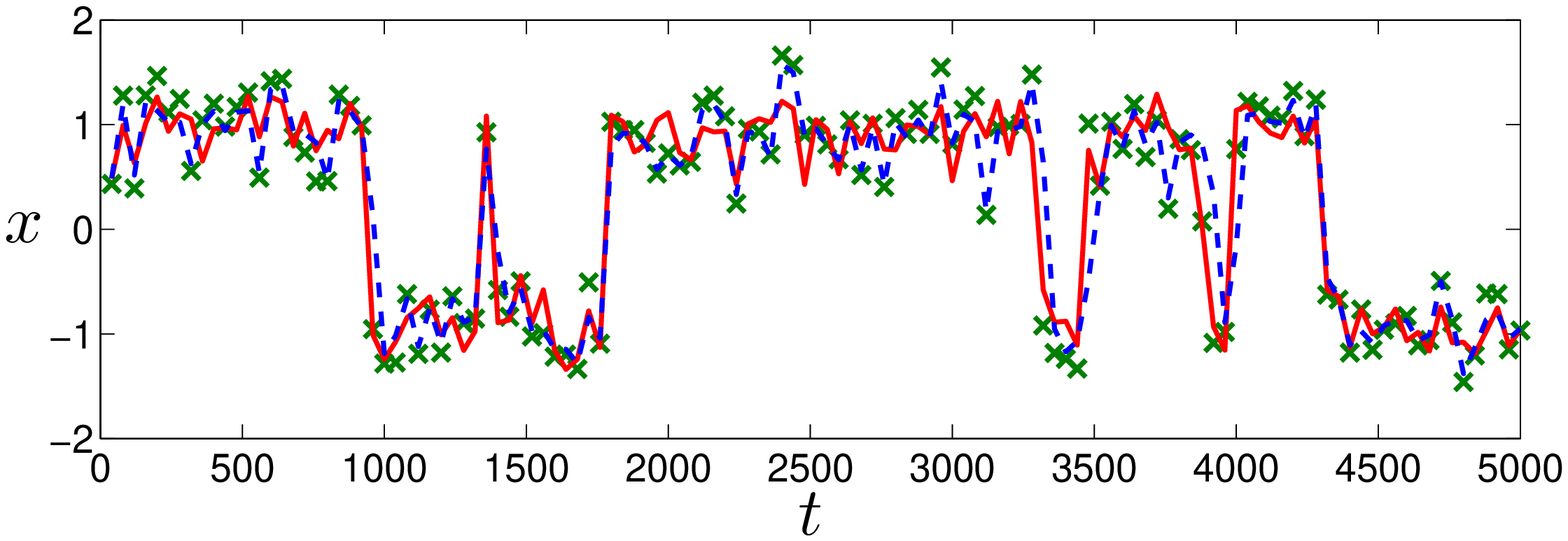}
\includegraphics[width=\columnwidth]{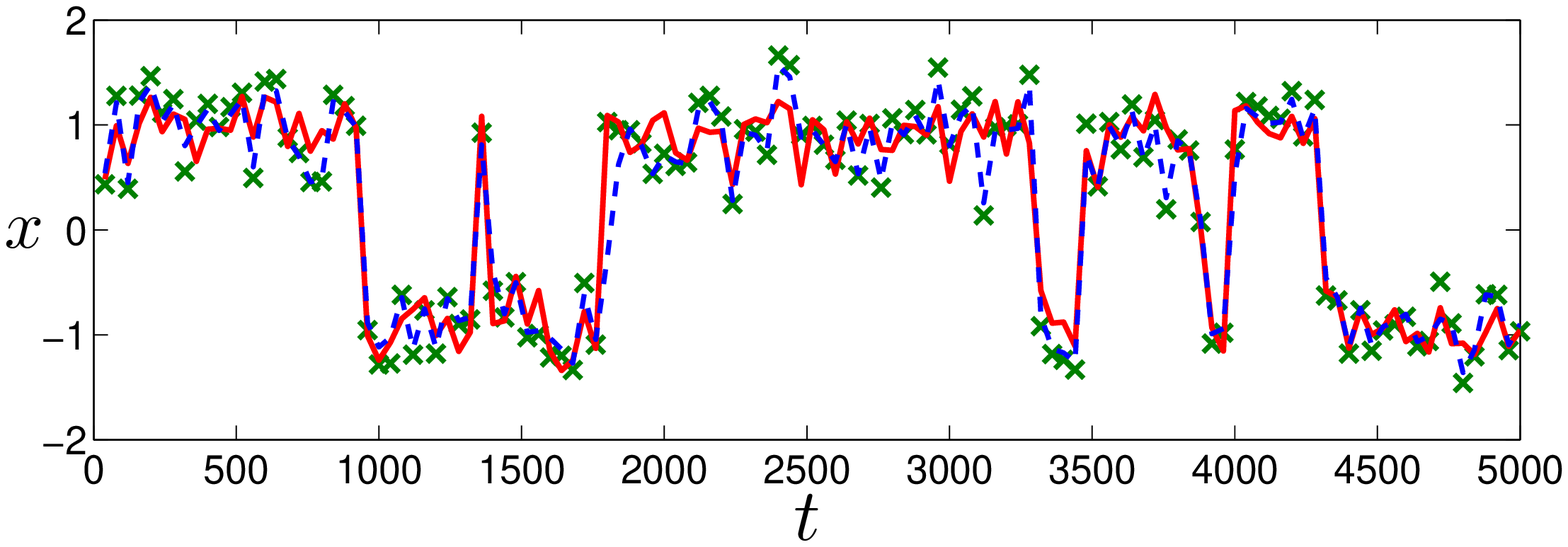}
\caption{Sample truth (solid line) and ETKF analyses (dashed line) for the observed $x$-component using the full deterministic 4D model (\ref{eqn:x})-(\ref{eqn:y3}) (top panel) and the reduced stochastic 1D climate model (\ref{eqn:climate}) (bottom panel) as the forecast model. The crosses are observations with error variance $R_{\rm obs} = 0.5 \sigma^2$ ($\sigma^2=0.126$). Here we have used an observation interval $\Delta \tobs = 50$ and used $k=15$ ensemble members.}
\label{fig:example}
\end{center}
\end{figure}

\noindent
Figure \ref{fig:example} shows two sample analysis time series created by assimilating the same truth and observations at regularly spaced intervals of $\Delta \tobs = 50$ where the stochastic 1D climate model and the full deterministic 4D model are used as the respective forecast models. Both forecast models track the truth well. The stochastic 1D climate model however tracks the transitions between the slow metastable states around $x^* = \pm1$ more accurately.

%

\begin{figure}[htbp]
\begin{center}
\includegraphics[width=0.7\columnwidth]{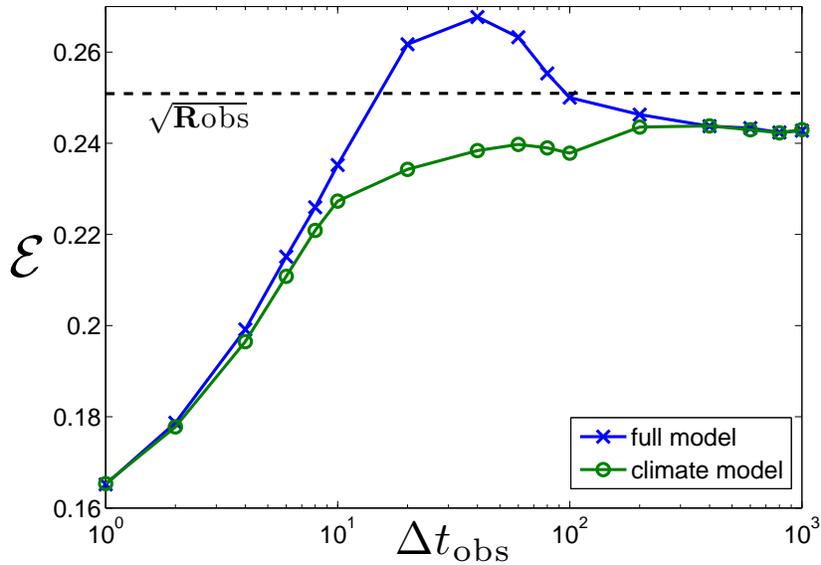}
\caption{RMS errors $\cal{E}$ of the analysis using the full deterministic 4D model (\ref{eqn:x})-(\ref{eqn:y3}) (crosses) and the stochastic 1D climate model (\ref{eqn:climate}) with $\sigma^2=0.126$ (circles) as forecast models, as a function of the observation interval $\Delta \tobs$. The horizontal dashed line indicates the observational error, $\sqrt{R_{\rm obs}} = \sigma/\sqrt{2} = 0.251$ ($\sigma^2=0.126$). The results are averaged over 1000 realisations.}
\label{fig:RMS_climate}
\end{center}
\end{figure}

\noindent
We quantify the improvement made using the reduced stochastic 1D climate model by measuring the RMS error $\cal E$  between the observed truth $x_t$ and the analysis mean $\bar{x}_a$,
\[
{\cal E} = \sqrt{\frac{1}{N} \sum_{i = 1}^N ( \bar{x}_a(t_i) - x_t(t_i))^2},
\]
where $N$ is the total number of analysis cycles. In Figure \ref{fig:RMS_climate} we show $\cal E$ as averaged over 1000 realizations, where we have used both the full deterministic 4D model and the reduced stochastic 1D climate model as forecast models, as a function of the observation interval $\Delta \tobs$. Both filters exhibit the same accuracy for small observation intervals $\Delta \tobs < 10$, when the system is frequently observed, and for very large observational times $\Delta \tobs > 500$, when the forecast is much larger than all the characteristic time
scales we discussed in Section~\ref{sec:timescales}. At very large observation intervals (not shown), when the ensemble of both forecast models will have explored the state space of the slow variable with climatological variance $\sigma_{\rm clim}^2 \approx 1$, the resulting analysis error covariance can be estimated using the Kirchoff-type addition rule of covariances (\ref{e.kirchoff}) as $P_a^{-1} = \sigma_{\rm clim}^{-2} + R_{\rm obs}^{-1}$ as $P_a=0.244^2$ for $R_{\rm obs} = 0.5\sigma^2$ and $\sigma^2 = 0.126$, which corresponds well with the measured asymptote of the RMS error ${\cal{E}}$ in Figure~\ref{fig:RMS_climate}. Interestingly, for moderate observation intervals $\Delta \tobs$ the stochastic 1D climate model performs considerably better than the full deterministic 4D model. Moreover, the analysis using the full deterministic 4D forecast model exhibits RMS errors worse than the observational error of $\sqrt{R_{\rm obs}} = 0.251$. In contrast, when the stochastic 1D climate model is used as a forecast model for the filter, the analysis error is always below the observational error, i.e. performing data assimilation is desirable over just trusting the observations (albeit with an improvement in RMS error of only around 5\%).\\

\begin{figure}[htbp]
\begin{center}
\includegraphics[width=0.7\columnwidth]{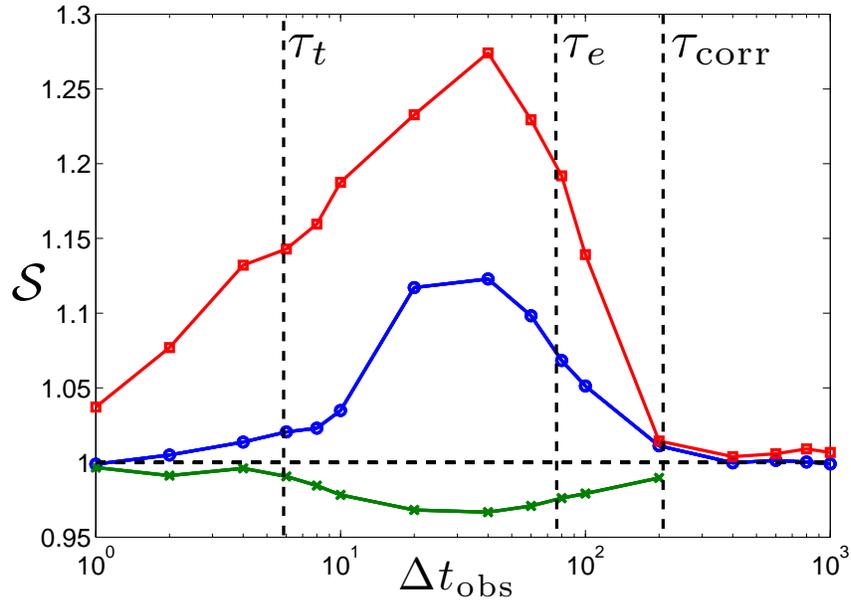}
\caption{Proportional skill improvement $\cal S$ of the stochastic 1D climate model (\ref{eqn:climate}) over the full deterministic 4D model (\ref{eqn:x})-(\ref{eqn:y3}) as a function of the observation interval $\Delta \tobs$. We show ${\cal S}$ as calculated over all analyses (circles), the metastable regimes only (crosses), and the transitions only (squares). Dashed vertical lines indicate the characteristic time scales of the slow dynamics: the decorrelation time $\tau_{\rm corr}$, first exit time $\tau_e$ and transit time $\tau_t$. Parameters are as in Figure \ref{fig:RMS_climate}.}
\label{fig:skill_climate}
\end{center}
\end{figure}

\noindent
We introduce the proportional improvement in the RMS error $\cal{E}$ or skill
\[
\cal S = \frac{{\cal E}_{\rm full}}{{\cal E}_{\rm clim}}\, ,
\]
where ${\cal E}_{\rm full}$ and ${\cal E}_{\rm clim}$ are the RMS errors when the full deterministic or the reduced stochastic model is used as forecast model, respectively. Values of ${\cal S} \geq 1$ imply that it is beneficial to use the reduced stochastic 1D climate model (note that when ${\cal S} = 1$ it is still beneficial to use the stochastic 1D climate model because of the reduced computational expense). Figure~\ref{fig:skill_climate} shows ${\cal S}$ as a function of the observation interval $\Delta \tobs$, averaged again over 1000 realizations, with the time scales $\tau_t$, $\tau_e$ and $\tau_{\rm corr}$ superimposed. Skill improvements occur only for observation intervals larger than the mean transit time $\tau_t$, with the maximum skill improvement over all forecasts occurring for $\Delta \tobs$ slightly smaller than the mean exit time $\tau_e$. If forecasts are longer than $\tau_{\rm corr}$, the full and climate models are essentially indistinguishable with ${\cal{S}}=1$.\\ Figure~\ref{fig:skill_climate} also shows how the skill ${\cal S}$ is distributed over the whole set of the analyses including metastable states and transitions between them, as well as only over those analyses which remain within a metastable state near either $x^\star = -1$ or $x^\star=1$, and those which fall in the transitions between the metastable states. Note that for large observation intervals $\Delta \tobs >\tau_{\rm corr}$ half of all forecasts switch between regimes, so the notion of analyses which remain within one metastable regime becomes obsolete. As such, we do not plot skill as calculated over the metastable regimes for $\tau > \tau_{\rm corr}$ in this case.\\ Figure~\ref{fig:skill_climate} illustrates that the major contribution to the skill improvement is from the stochastic 1D climate model better detecting the transitions between slow metastable states. While there is actually a minor degradation in skill of around $3\%$ for forecasts with $\tau_t < \Delta \tobs < \tau_{\rm corr}$ made for analyses which remain within metastable states, skill improvements of over $25 \%$ are made for those analyses which fall into the transitions between the metastable states. This is because analyses which detect transitions between metastable states incorrectly contribute large RMS errors (${\cal O}(1)$), compared to the small errors (${\cal O}(R_{\rm obs})$) made by incorrect analyses within the metastable states. These large errors have a major impact upon the total skill $\cal S$.\\ 
There is also skill improvement as calculated over the transition periods for small values of the observation interval $\Delta \tobs$. However, there is no corresponding overall skill improvement since for small observation intervals the number of instances where the truth remains within a metastable regime is overwhelmingly larger, thereby swamping the skill improvement obtained over the rare transitions.\\

\noindent
We note that the skill $\cal S$ is insensitive to the observational noise level, provided $R_{\rm obs}$ is sufficiently small so that observations do not misrepresent the actual metastable state or regime in which the truth resides. The skill $\cal S$ is roughly the same for $R_{\rm obs}=0.1 \sigma^2$ and $R_{\rm obs}=0.25 \sigma^2$, and we observe no skill improvement ${\cal S} \approx 1$ when we take very poor observations with $R_{\rm obs} = 0.5$, approximately half the climatological variance of the full slow system.


\subsection{Sensitivity of analysis skill to uncertainties in the diffusion coefficient}

\noindent
As for the time scales in Section \ref{sec:sensitivity}, we examine how the analysis skill $\cal S$ depends on the diffusion coefficient  $\sigma^2$ (see the Appendix for dependence on the drift term). Figure \ref{fig:skill_sigma} shows the skill ${\cal{S}}$ as a function of the observation interval $\Delta \tobs$ for several values of $\sigma^2$. For very small diffusion and for very large diffusion (not shown) the skill is actually smaller than $1$ for all observation intervals, implying that the stochastic 1D climate model performs worse than the full deterministic 4D model. There exists a range of values of $\sigma^2$ for which the stochastic climate model outperforms the full deterministic model with roughly similar skill values ${\cal S}(\Delta \tobs)>1$ for $\tau_t < \Delta \tobs < \tau_{\rm corr}$.\\ The dependence of analysis skill upon the diffusion coefficient can be understood as follows: for small diffusion ($\sigma^2 = 0.1$) the stochastic 1D climate model is less diffusive than the truth and the deterministic 4D forecast model, and its associated forecast is likely to remain in a  metastable state. This produces small ensemble spread and therefore causes more instances of the forecast not detecting a transition between regimes. On the other extreme, for large diffusion $\sigma^2 \gg 0.15$, the stochastic 1D climate model exhibits regime switches too frequently, thereby producing a large forecast error which contaminates the analysis. For moderate values of the diffusion coefficient, the increase in spread of the forecast ensemble helps the detection of regime switches. 

%
\begin{figure}[htbp]
\begin{center}
\includegraphics[width=0.7\columnwidth]{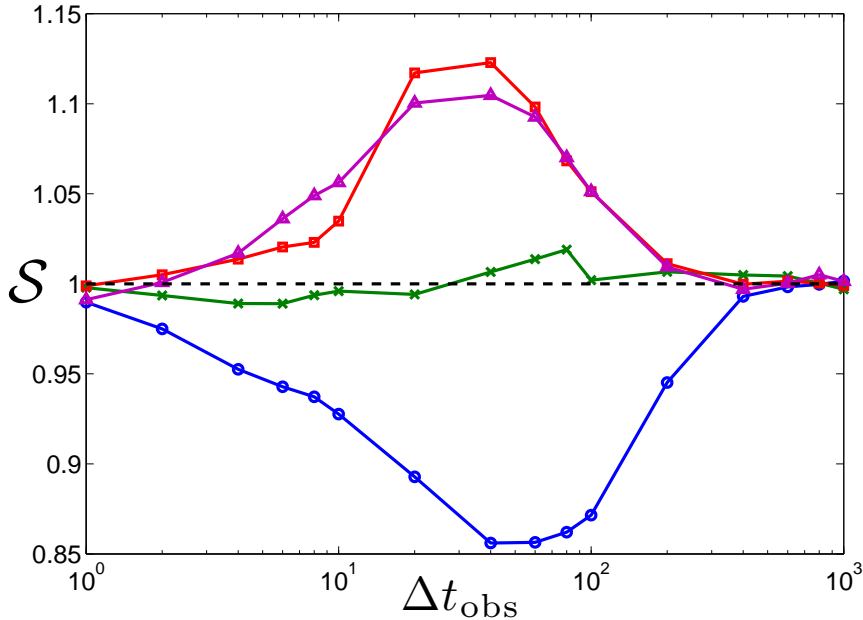}
\caption{Proportional skill improvement $\cal S$ of the stochastic 1D climate model (\ref{eqn:climate}) over the full deterministic 4D model (\ref{eqn:x})-(\ref{eqn:y3}) as a function of the observation interval $\Delta \tobs$ with $\sigma^2 = 0.1$ (circles), $\sigma^2 = 0.113$ (crosses), $\sigma^2 = 0.126$ (squares) and $\sigma^2 = 0.15$ (triangles). All other parameters are as in Figure \ref{fig:RMS_climate}.}
\label{fig:skill_sigma}
\end{center}
\end{figure}
%


\subsection{Effect of ensemble size and covariance inflation}
The results from the previous subsection suggest that the observed skill improvement is linked to the increased forecast ensemble spread of the stochastic climate model, rather than to the accuracy of the stochastic parametrization. This is exemplified by the fact that for $\sigma^2=0.113$ for which the stochastic 1D climate model approximates the statistics of the full deterministic 4D model best, there is no skill improvement ${\cal S}\approx 1$, and skill improvement ${\cal S} >1$ is given for $\sigma^2>0.113$ where transitions between the metastable states are better captured.\\

\noindent
The lack of sufficient ensemble spread when using the full deterministic system as a forecast model is caused by the finite size of the ensemble \citep{Ehrendorfer07}. In the data assimilation community a common approach to account for insufficient spread is covariance inflation \citep{AndersonAnderson99}, whereby the forecast variance $\P_f$ is artificially inflated by multiplication with a factor $\delta>1$. We now investigate how the ensemble size $k$ and inflation factor $\delta$ affect the skill. In particular we show that the RMS error ${\mathcal{E}}$ using the full deterministic model can be reduced by either increasing the ensemble size $k$ or by using larger inflation factors $\delta$.\\

\begin{figure}[htbp]
\begin{center}
\includegraphics[width=0.7\columnwidth]{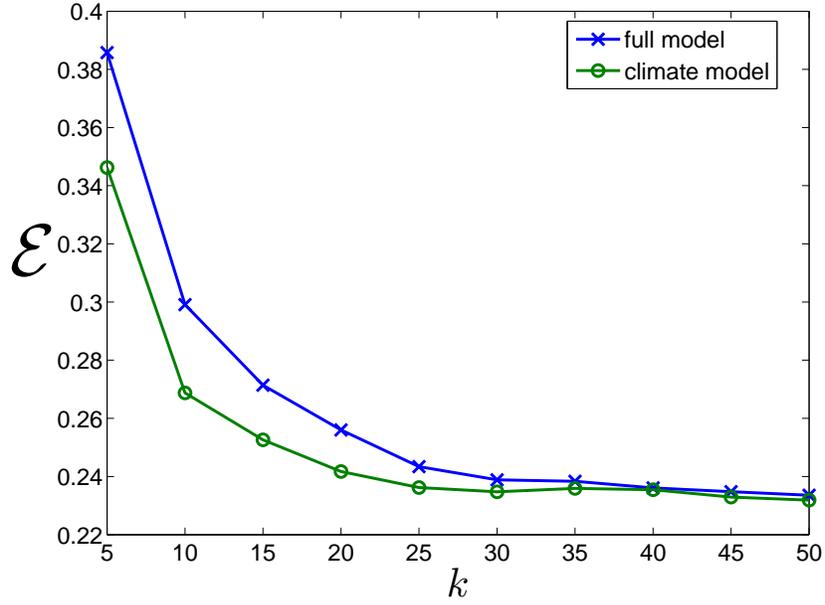}
\caption{RMS error ${\cal E}$ as a function of ensemble size $k$, using the full deterministic 4D model (\ref{eqn:x})-(\ref{eqn:y3}) (crosses) and the stochastic 1D climate model (\ref{eqn:climate}) with $\sigma^2 = 0.126$ (circles) as forecast models. We used $\Delta \tobs = 40$; all other parameters as in Figure \ref{fig:RMS_climate}.}
\label{fig:RMS_Nens}
\end{center}
\end{figure}

\noindent
Figure \ref{fig:RMS_Nens} shows $\cal E$ as a function of the ensemble size $k$ for $\Delta \tobs = 40$ (which is close to the observation interval yielding maximal skill improvement ${\cal S}$; see Figure~\ref{fig:skill_climate}), $\sigma^2 = 0.126$, and averaged over 200 realizations with $\sigma^2=0.126$. For large ensembles with $k \geq 35$ the overdiffusive stochastic 1D climate model and the full deterministic 4D model perform equally well as forecast models. We have checked that when using a large ensemble ($k = 50$) the stochastic climate and full deterministic models perform equally well for the range of observation intervals used in Figure~\ref{fig:RMS_climate}. For smaller ensemble sizes $k$ the stochastic 1D climate model outperforms the full deterministic 4D model. This illustrates that the skill improvement is linked to the insufficient ensemble spread of the deterministic 4D model which is compensated in the stochastic 1D climate model by a slightly increased diffusion coefficient.\\

\begin{figure}[htbp]
\begin{center}
\includegraphics[width=0.7\columnwidth]{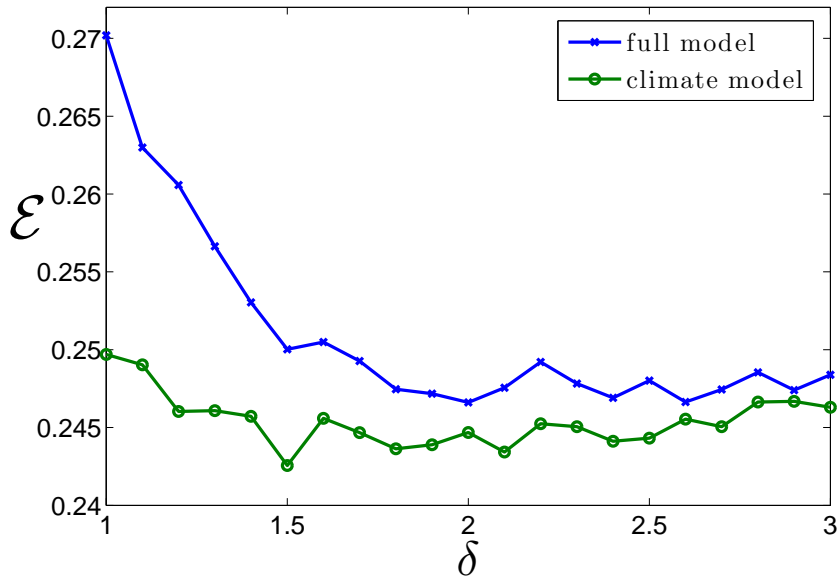}
\caption{RMS error ${\cal E}$ as a function of inflation factor $\delta$, using the full deterministic 4D model (\ref{eqn:x})-(\ref{eqn:y3}) (crosses) and the stochastic 1D climate model (\ref{eqn:climate}) with $\sigma^2 = 0.126$ (circles) as forecast models. We used $\Delta \tobs = 40$; all other parameters are as in Figure \ref{fig:RMS_climate}.}
\label{fig:RMS_inf}
\end{center}
\end{figure}

\noindent
Similarly, for fixed ensemble size $k$, the ensemble spread can be artificially increased by inflating the forecast covariance with a constant factor $\delta$. Figure \ref{fig:RMS_inf} shows $\cal E$ as a function of the inflation factor $\delta$, for $1 \leq \delta \leq 3$. It is seen that for unrealistically large values of $\delta$ the RMS error decreases when the full deterministic 4D model is used as a forecast model, whereas it is relatively insensitive to changes in $\delta$ for the stochastic 1D climate model. For even larger values of $\delta$ the RMS errors for the two forecast models eventually become equal (not shown).

%


\subsection{Reliability and ranked probability diagrams}

\noindent
Another way of understanding why the ensemble variance of the stochastic 1D climate model produces skill better than that of the full deterministic 4D model is through ranked probability histograms (also known as Talagrand diagrams) \citep{Anderson96,Hamill97,Talagrand97}. To create probability density histograms we sort the forecast ensemble $\X_f = [x_{f,1},x_{f,2}, ... , x_{f,k}]$ and create bins $(-\infty,x_{f,1}]$, $(x_{f,1},x_{f,2}]$, ... , $(x_{f,k},\infty)$ at each forecast step. We then increment whichever bin the truth falls into at each forecast step to produce a histogram of probabilities $P_i$ of the truth being in bin $i$. 
A reliable ensemble is considered to be one where the truth and the ensemble members can be viewed as drawn from the same distribution. A flat probability density histogram therefore is seen as indicating a reliable ensemble for which each ensemble member has equal probability of being nearest to the truth. A convex probability density histogram indicates a lack of spread of the ensemble, while a concave diagram indicates an ensemble which exhibits too large spread. We direct the reader to \cite{Wilks} for a detailed discussion.\\

\noindent
\cite{Hamill01a} suggested that a single ranked probability histogram is not sufficient for dynamical systems with several dynamical regimes as in our case, but that in fact one must look at the variability of the ensemble in the different dynamical regions individually. We therefore construct ranked probability histograms (for both forecast models) over the metastable regime with characteristic time $\bar \tau$ only and over the complementary regime of transitions with characteristic time $\tau_t$ only, as done for the distribution of the skill over those regimes (see Figure~\ref{fig:skill_climate}).\\

\begin{figure}[htbp]
\begin{center}
\includegraphics[width=0.55\columnwidth]{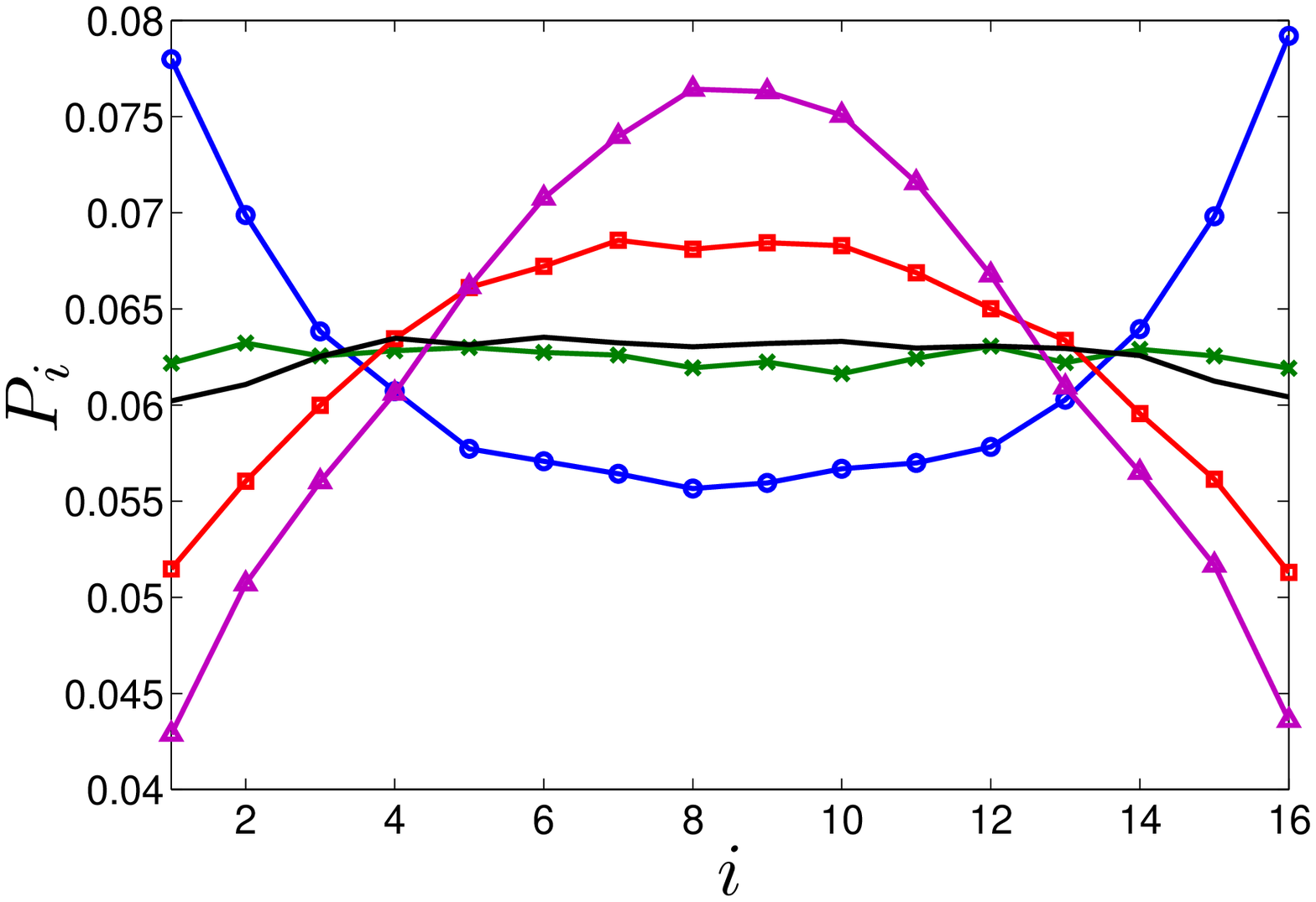}
\includegraphics[width=0.55\columnwidth]{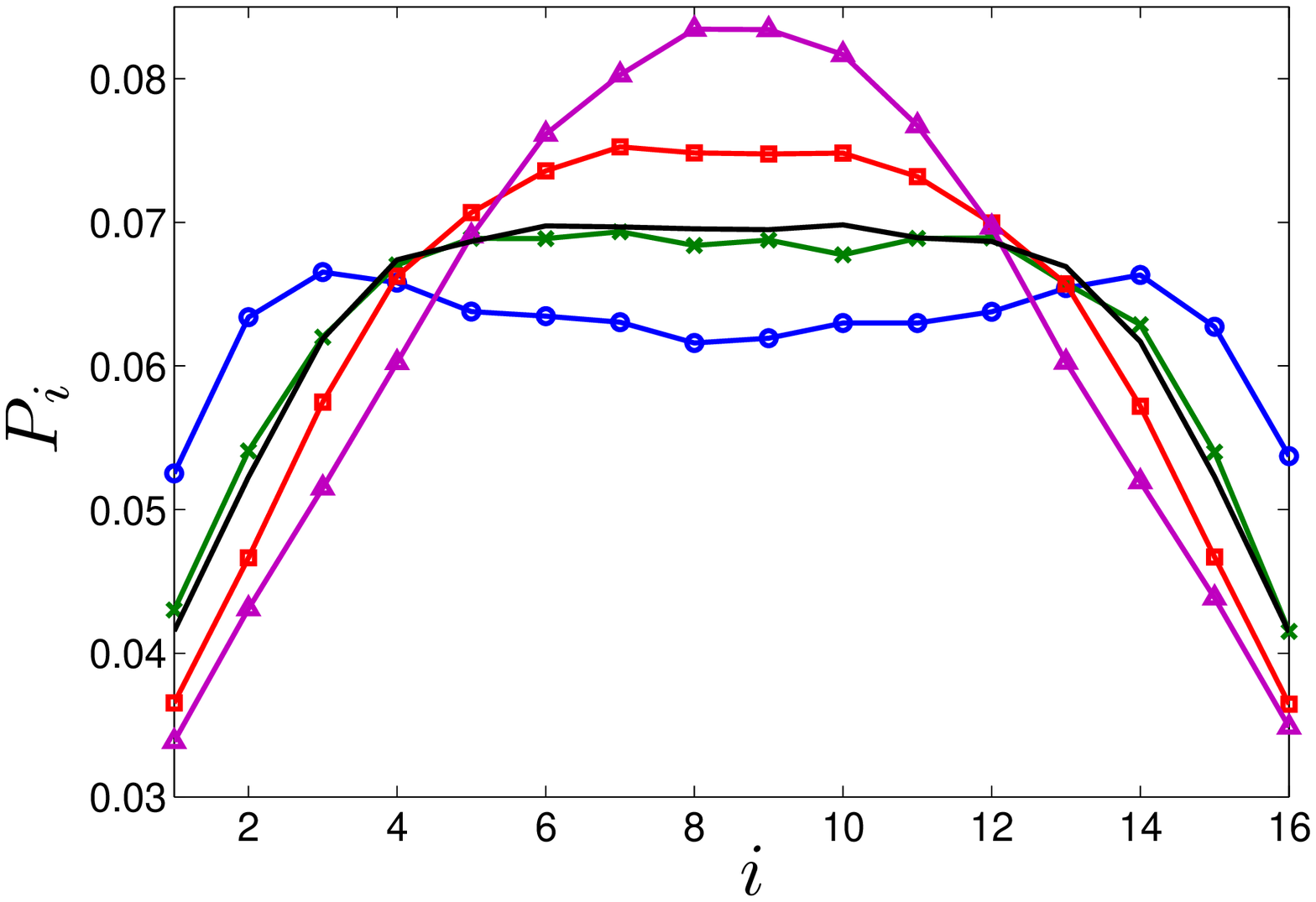}
\includegraphics[width=0.55\columnwidth]{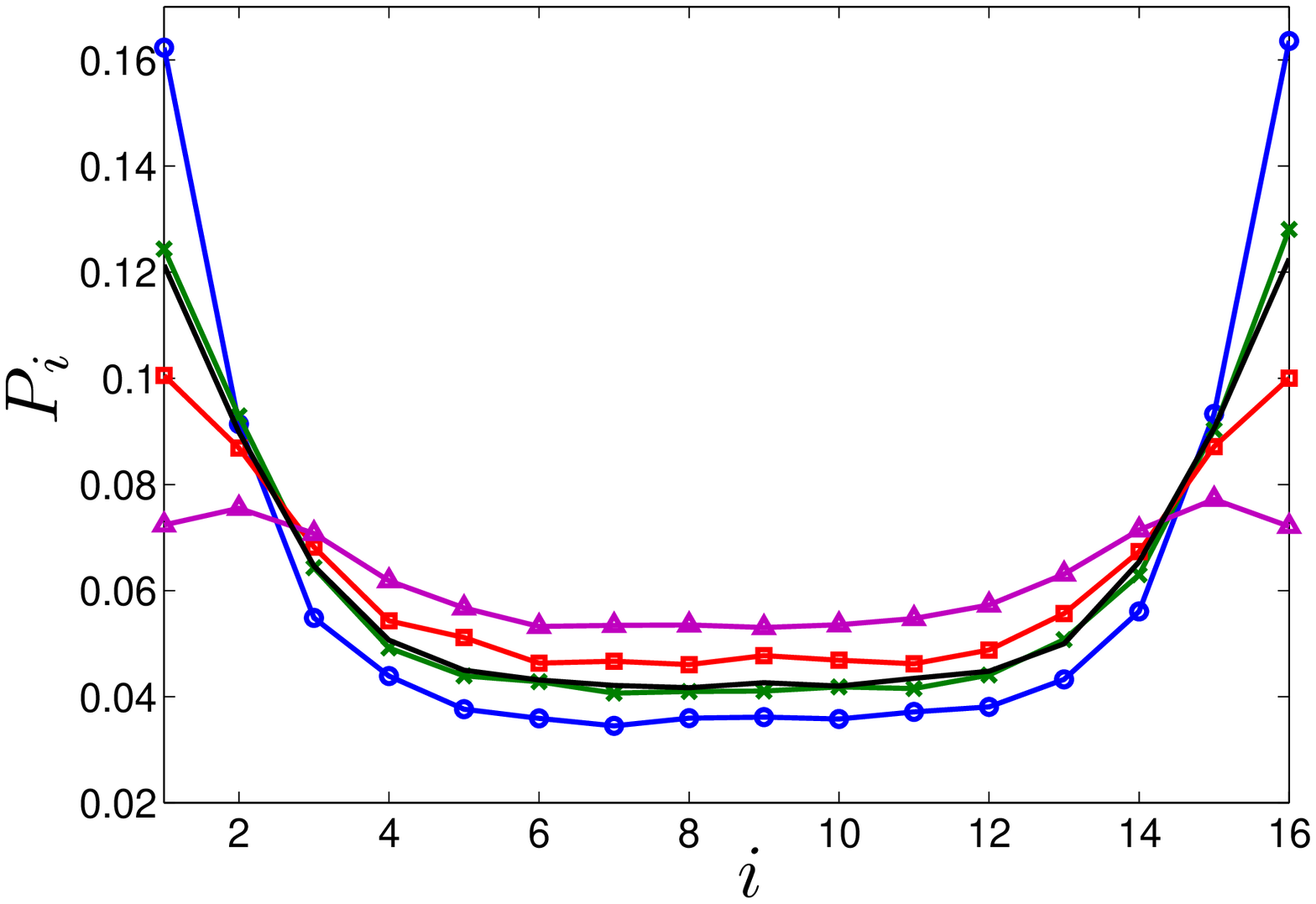}
\caption{Ranked probability histograms for the full deterministic 4D model (\ref{eqn:x})-(\ref{eqn:y3}) (dashed line) and the stochastic 1D climate model (\ref{eqn:climate}) with $\sigma^2 = 0.1$ (circles), $\sigma^2 = 0.113$ (crosses), $\sigma^2 = 0.126$ (squares) and $\sigma^2 = 0.15$ (triangles) over all forecasts (a), as well as only those forecasts which fall near the metastable states $x^* = \pm 1$ (b) and those which fall in the transitions between metastable states (c).}
\label{fig:talagrand}
\end{center}
\end{figure}

\noindent
In Figure \ref{fig:talagrand}, we plot probability density histograms of the forecast for the full deterministic 4D model (\ref{eqn:x})-(\ref{eqn:y3}) and the stochastic 1D climate model (\ref{eqn:climate}) with various values of the diffusion coefficient $\sigma^2$, (a) over all analyses, (b) over the analyses which fall near the metastable states $x^* = \pm 1$ and (c) over the analyses which fall in the transitions between metastable states. We use a long analysis cycle containing $250,000$ forecasts. Although the full deterministic model produces the most reliable ensemble when averaged over all forecasts, it overestimates the variance in the metastable regions and underestimates the variance across the transitions. For the stochastic 1D climate model, the more underdiffusive model with $\sigma^2 = 0.1$ produces the most reliable ensemble in the metastable regions while the more diffusive models with $\sigma^2 = 0.126$ and $\sigma^2 = 0.15$ both produce more uniform ranked probability histograms than the full model does in the transition regions. The reduced stochastic 1D climate model with optimal diffusion coefficient $\sigma^2=0.113$ behaves similarly to the full deterministic 4D model as expected. It is thus more desirable to favour the more diffusive stochastic climate models because a comparatively large improvement in RMS error can be made in the transition regions compared to in the metastable regions. This agrees with our earlier observation that the skill improvement is entirely due to the stochastic climate model being more accurate in the transition regions where large errors can be accrued by misclassifying the regime (see Figure~\ref{fig:skill_climate}).\\

%


\section{Discussion}
\label{sec:discussion}

\noindent
We have investigated a homogenized stochastic climate model of a bistable multiscale system with a chaotic fast subsystem, and its usage as a forecast model in ensemble Kalman filtering. The reduced stochastic climate model was numerically shown to faithfully reproduce the statistics of the slow dynamics of the full deterministic model even in the case of finite time scale separation where the theorems underpinning homogenization fail. Homogenized stochastic climate models replace in a controlled fashion the accumulative effect of the fast chaotic dynamics on the slow dynamics by white noise. We estimated the analytical expressions of the drift and diffusion terms of the reduced stochastic model using coarse-graining and quadratic variations, and found that the diffusion coefficient is very sensitive to the applied undersampling time. The statistics of the stochastic climate model, i.e. the probability density function and averaged characteristic time scales, were found to be very sensitive to the diffusion and drift terms used. Despite this sensitivity and the implied uncertainties associated with the ``correct'' stochastic climate model, we found that stochastic climate models can be beneficial as forecast models for data assimilation in an ensemble filter setting. We have shown numerically that using such a stochastic climate model has the advantage of 1.) being computationally more efficient to run due to the reduced dimensionality and avoidance of stiff dynamics, and 2.) producing a superior analysis mean for a range of observation intervals $\Delta \tobs$. These skill improvements occur for observation intervals larger than the time taken to switch between slow metastable states $\tau_t$ but less than the decorrelation time $\tau_{\rm corr}$.\\

\noindent
This skill improvement is due to the associated larger ensemble variance of the stochastic climate model, which allows the ensemble to explore the full state space of the slow variable better than the full deterministic model does. Forecasts made by the stochastic climate model will generally contain more ensemble members which cross the potential barrier between the slow metastable states (weather or climate regimes) than full model forecasts will. Consequently, there is a greater chance during the forecast of one or more climate model ensemble members crossing the potential barrier to the opposite metastable state than in the full deterministic model ensemble. If an observation appears near the other metastable state from where the forecast began, the Kalman filter analysis equation (\ref{eqn:zaens}) trusts the observation more than the forecast, correctly producing an analysis which changes between the regimes. Thus the stochastic climate model with large enough diffusion better detects transitions between slow metastable states and so has lower RMS error near these transitions in which large errors can be accrued. We remark that strict time scale separation is not necessary for the existence of metastable regimes; we have checked that for $\epsilon=1$ similar dynamics as depicted in Figure~\ref{fig:taui} can be observed for increased coupling strength. In this case we can still obtain skill for observation intervals between the transit time and the decorrelation time when using a stochastic Langevin equation of the form (\ref{eqn:climate}), albeit with decreased diffusion whose dynamically optimal value is no longer approximated by homogenization.\\

Similarly, we remark that multimodal probability density functions are not necessary for the existence of metastable regimes, as was pointed out by \cite{Wirth01,Majda06}. In cyclostationary systems, i.e. systems which involve time-periodic coefficients, or in aperiodic systems in which ``regimes" occur intermittently, the probability density function may be unimodal whereas if restricted to time windows focussing on the regimes the ``hidden" regimes can be identified. This has important implications for interpreting atmospheric data \citep{Majda06,Franzke08,Franzke09} where one needs to reconcile the apparent paradox of multimodal probability density functions of planetary waves obtained from (relatively short-time) observational data \citep{KimotoGhil93,ChengWallace93,CortiEtAl99} and unimodal but non-Gaussian probability density functions obtained from long-time simulations of atmospheric general circulation models \citep{Kondrashov04,BranstatorBerner05,FranzkeMajda06,BernerBranstator07} (for a critical account on the analysis of short climatic data see for example \cite{HsuZwiers01,Stephenson04,Ambaum08}).\\ 

However, it is pertinent to mention that the results found here are not dependent on the existence of metastable states, but only require dynamics with rapid large amplitude excursions as for example in intermittent systems. 
%
%
%
In \cite{HarlimMajda08,HarlimMajda10} a system without multiple equilibria was investigated and superior filter performance was found for stochastic parametrizations (in this case by radically replacing all nonlinear terms by linear Ornstein-Uhlenbeck processes in Fourier space). We believe that their observed increased analysis skill can be explained by the increased ensemble spread counteracting finite sampling errors of the underlying ensemble Kalman filter approach as we did. The advantage when homogenization can be used is that the slow dynamics and its mean are still well resolved and reproduced and therefore the ensemble spread increasing stochasticity is of a more controlled nature. The use of homogenized stochastic climate models with adjusted larger diffusivity therefore amounts to incorporating deliberate but controlled model error into the data assimilation scheme.\\

\noindent
We found that the same skill can be obtained using the full deterministic model either by using a larger ensemble, or by increasing the covariance inflation factor. The first is undesirable for large models because of the computational difficulty of simulating large ensembles and their covariances, while the second is undesirable because it introduces unphysical inflation (here a multiplicative factor of $\delta > 2$ was needed) and requires expensive empirical tuning of the inflation factor.\\

\noindent
Using stochastic reduced models for multi-scale systems has computational advantages. First of all, the dimensional reduction allows for a a gain in computational speed. Furthermore, by replacing a stiff ordinary differential equation by a non-stiff stochastic differential equation means that coarser integration time steps can be used which otherwise would lead to a break down when simulating the stiff multi-scale system.\\

\noindent
Homogenized stochastic climate models therefore provide a cheap way of simulating large ensembles, and a more natural way of incorporating forecast covariance inflation into the data assimilation algorithm. In particular, we find that the stochastic climate model produces a more reliable ensemble (as characterised by the ranked probability histogram) which better detects the transitions between slow metastable states. This leads to a far superior analysis in these regimes than that produced by the full deterministic model.

\section*{Acknowledgments}
GAG would like to thank Jochen Br\"ocker and Balu Nadiga for bringing ranked probability diagrams to our attention. LM acknowledges the support of an Australian Postgraduate Award. GAG acknowledges support from the Australian Research Council.


\begin{appendix}
\section*{Appendix: Sensitivity of the stochastic climate model to uncertainties in the drift term}
\label{sec:appendix}
In this appendix we examine how the statistics of the stochastic 1D climate model (\ref{eqn:climate}) depends on uncertainties in the estimation of the drift term
\[
d(X) = a x(b-x^2)\; .
\]
The parameter $b$ controls the location of the metastable states near $x^* = \pm \sqrt{b}$ and their separation. The height of the potential barrier $\Delta V(x)=ab^2/4$ is controlled by $a$ and $b$.\\

\begin{figure}[htbp]
\centering
\includegraphics[width = 0.7\columnwidth]{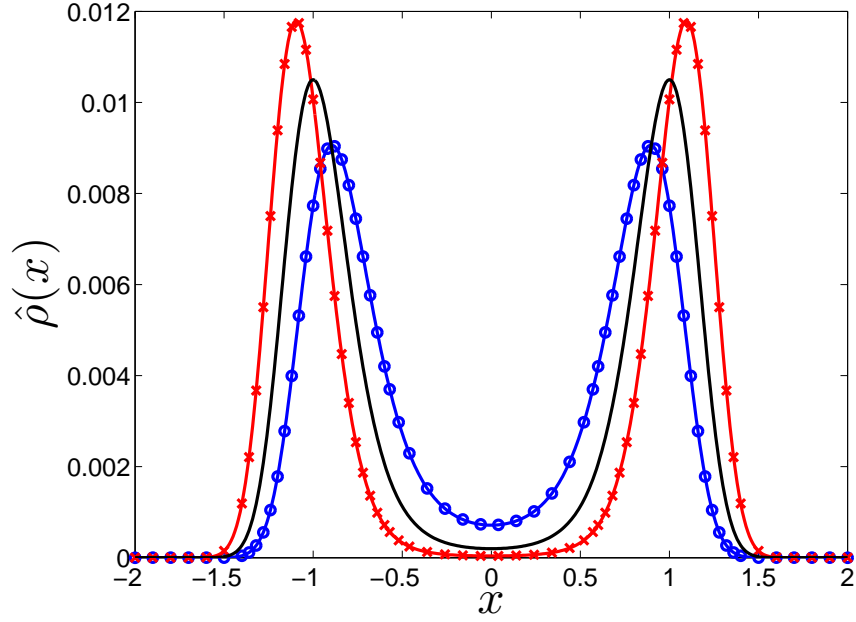}
\caption{Invariant density (\ref{e.rhohat}) for the stochastic 1D climate model (\ref{eqn:climate}), with $\sigma^2=0.126$ and $(a,b)=(1,1)$ (solid line), $(a,b) = (1,0.8)$ (circles) and $(a,b) = (1,1.2)$ (crosses).}
\label{f.pdf_drift}
\end{figure}

\noindent
In Figure \ref{f.pdf_drift} we show the invariant density (\ref{e.rhohat}) for the stochastic 1D climate model with $\sigma^2 = 0.126$ and several combinations of the drift term parameters $a$ and $b$. The invariant density is more sensitive to changes in $b$ than in $a$ (not shown) since $b$ simultaneously affects the distance of the minima of the potential and the height of the potential barrier. The location of the maxima of the probability density function changes by approximately $22 \%$ and the actual values of the maxima change by approximately $27 \%$ over the range of $b$ values shown. Varying $a$ produces similar changes in $\hat{\rho}(x)$, but as expected the locations of the maxima of the probability density function are not shifted.\\

\noindent
As for the diffusion coefficient, differences in the approximation of the empirical density imply a sensitivity of the statistics to changes in the drift term parameters. In Table~\ref{tbl:timescales_drift} we show how the characteristic time scales change for varying drift term parameters. Note that now the transit time $\tau_t$ varies because by varying $a$ we modify the rate of divergence of the unstable fixpoint.\\
%
\begin{table}[htbp]
\caption{Characteristic timescales of the stochastic 1D climate model (\ref{eqn:climate}) with $\sigma^2 = 0.126$ for different values of the drift coefficients $a$ and $b$.}\label{tbl:timescales_drift}
\begin{center}
\begin{tabular}{|c|c|c|c|c|}
\hline
 &  $a=0.8$ & $a=1.2$ & $a=1$ & $a=1$\\
&  $b=1$ & $b=1$ & $b=0.8$ & $b=1.2$\\
\hline
$\tau_{\rm corr}$ &   77.3  & 256.3  & 41.1 & 645.7  \\
$\tau_e$ & 43.7 & 136.1 & 31.1 & 212.9 \\
$\tau_t$ &  6.4  & 4.8 & 7.2 & 4.5 \\
\hline
\end{tabular}
\end{center}
\end{table}

\begin{figure}[htbp]
\begin{center}
\includegraphics[width=0.7\columnwidth]{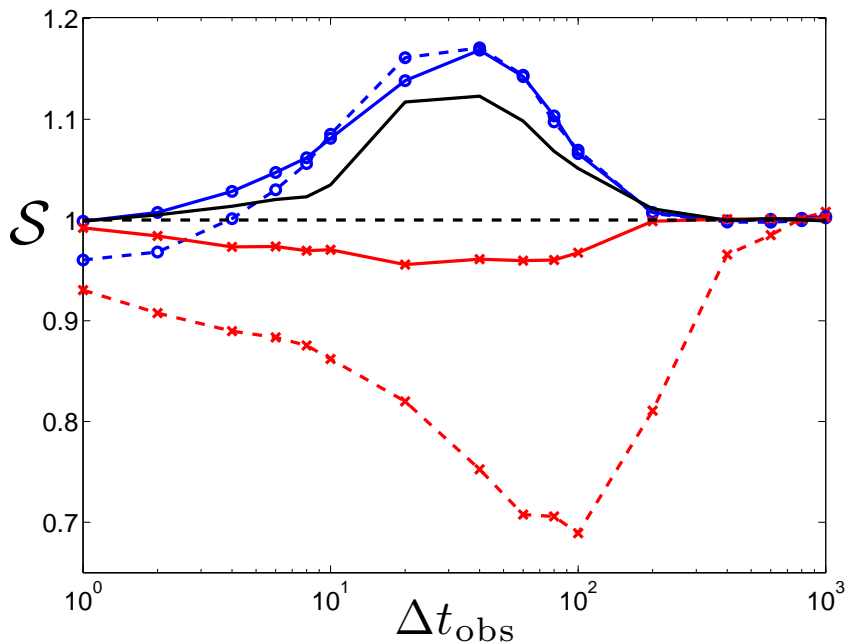}
\caption{Proportional skill improvement $\cal S$ of the stochastic 1D climate model (\ref{eqn:climate}) over the full deterministic 4D model (\ref{eqn:x})-(\ref{eqn:y3}) as a function of the observation interval $\Delta \tobs$ for $\sigma^2 = 0.126$, with $(a,b)=(1,1)$ (black solid line), $(a,b) = (0.8,1)$ (solid, circles), $(a,b) =(1.2,1)$ (solid, crosses), $(a,b) = (1,0.8)$ (dashed, circles) and $(a,b) = (1,1.2)$ (dashed, crosses).  All other parameters are as in Figure \ref{fig:RMS_climate}.}
\label{fig:skill_drift}
\end{center}
\end{figure}

\noindent
We now investigate how uncertainties in the drift term parameters $a$ and $b$ may affect the skill ${\cal S}$ in an ETKF data assimilation procedure. Figure \ref{fig:skill_drift} shows skill curves for fixed value of the diffusion coefficient $\sigma^2 = 0.126$, for different combinations of $a$ and $b$. As with the characteristic time scales, the skill $\cal S$ is more sensitive to changes in $b$ than in $a$. The sensitivity to changes in the drift term is of a similar nature to the one for diffusion. Whereas decreasing $a$ or $b$ by $20 \%$ produces little change in skill, increasing $a$ or $b$ by $20\%$ makes the stochastic 1D climate model again less skilful than the full deterministic 4D model. This is readily explained by noticing that, increasing $b$ increases the difference between the two metastable states and also the height of the potential barrier; therefore, for fixed diffusion $\sigma^2$, this inhibits transitions between the metastable states, thereby decreasing the spread of the ensemble. Similarly, decreasing $a$ (while keeping $b$ fixed) increases the potential barrier, inhibiting transitions and thereby decreasing the ensemble spread.\\ As for the diffusion coefficient $\sigma^2$, the values of the parameters of the drift term which are closest to the actual values $a=b=1$ for which the stochastic climate model best approximates the slow dynamics of the full deterministic 4D model is not the optimal value in terms of skill improvement. Values which increase the ensemble spread are more favourable.

\end{appendix}


{\clearpage}
\bibliographystyle{ametsoc}

\bibliography{bibliography}

\end{document}